\documentclass[12pt,a4paper]{article}
\usepackage[ascii]{inputenc}
\usepackage{amsmath}
\usepackage{booktabs}
\usepackage{amsfonts}
\usepackage{amssymb}
\usepackage{makeidx}
\usepackage{graphicx}
\usepackage{algorithm}
\usepackage[noend]{algpseudocode}
\usepackage{algcompatible}
\usepackage{xcolor}
\usepackage{lineno}
\usepackage{authblk}
\usepackage{setspace}
\usepackage{hyperref}
 \hypersetup{
     colorlinks=true,
     linkcolor=blue,
     filecolor=blue,
     citecolor = blue,      
     urlcolor=cyan,
     }
\graphicspath{{newgraph/}}
\usepackage[flushleft]{threeparttable}
\usepackage{natbib}
\usepackage{subfig}
\usepackage{float}
\usepackage{graphicx}
\usepackage[width=18.00cm, height=25.00cm]{geometry}
\usepackage[flushleft]{threeparttable}
 \author[1*]{Ezra Gayawan}
 \affil[1]{\small Department of Statistics, Federal University of Technology, Akure, Nigeria 
 \thanks{Corresponding Author: egayawan@futa.edu.ng}}
 \author[2]{Osafu Augustine Egbon}
 \affil[2]{\small Institute of Mathematical and Computer Sciences, University of S\~ao Paulo, S\~ao Carlos, Brazil}
 \author[3]{Edson Utazi}
 \affil[3]{\small WorldPop, School of Geography and Environmental Science, University of Southampton, University
Rd, Southampton, SO17 1BJ, UK}
\author[4]{Jamila Abubakar Umar}
\affil[4]{\small 
National Primary Health Care Development Agency, Abuja, Nigeria}
\affil[5]{\small Medical Research Council Centre for Global Infectious Disease Analysis, 
Imperial College London, London, UK}
\author[5,6]{Caroline Trotter}
\affil[6]{\small Department of Veterinary Medicine and Pathology, University 
of Cambridge, Cambridge, UK}

\title{The impact of
Women's empowerment on childhood
vaccination coverage in Nigeria: a
spatio-temporal analysis}
\begin{document}
	\maketitle
		\section*{Abstract}
Immunization is one of the most impactful public health achievements, significantly reducing childhood morbidity and mortality worldwide. However, gender disparity and women's disempowerment constitute structural barriers in accessing vaccine services in low- and middle-income countries. In Nigeria, widespread differences in social norms and cultural values affect gender roles and influence women's ability to decide their own healthcare needs and participate in household decision-making. This leads to attitudinal differences in uptake of immunization depending on the child's location of residence. 
Using data from four waves of the Nigeria Demographic and Health Survey, we constructed two empowerment indices that determine whether caregivers participate in household decision-making 
and have the ability to 
decide on their healthcare needs. We used a structured spatiotemporal statistical model to 
determine whether a significant part of childhood vaccination coverage disparities can be attributed to these women's empowerment measures and predicted events at the third administrative level of the country. We considered five vaccination indicators: Bacillus Calmette-Guerin (BCG), zero-dose, receiving a complete dose of DPT, MCV-1 (first dose of measles containing vaccine), and receipt of all basic vaccinations. 
The adopted model was validated by comparing the empirical estimates of vaccination coverage level from the data with model projections at the second administrative level. 
The findings indicate that although empowerment regarding participation in household decision-making and healthcare utilization is generally associated with increased vaccine uptake, their effects vary considerably across locations and notably among the highly empowered category of women. 
 Although there are efforts to bridge immunization gaps within the country, the study emphasizes the need for tailored strategies that target up-scaling the ability of women and the wider community to participate in decision-making process and be able to decide on healthcare needs to address regional disparities and improve vaccination coverage.

\textbf{Keywords}: Immunization; Women's empowerment; Decision-making; Healthcare utilization; Structural model; Bayesian analysis

\section{Introduction}
Immunization is one of the most successful public health interventions that has greatly contributed to life-saving efforts. In recent decades, the availability and expansion of vaccination for children has contributed to attaining child health indicators through the reduction of illness, disability, and mortality from preventable diseases \cite{world2019global}. Notwithstanding, vaccine-preventable diseases are still a major cause of morbidity and mortality in many low- and middle-income countries (LMIC) and it has been challenging to reach the last 20\% of the children to receive vaccines protecting against measles, pneumococcal pneumonia, and rotavirus \cite{local2021mapping,mosser2019mapping}. The disruption caused by COVID-19 and the efforts towards COVID-19 vaccination also strained health systems leading to setbacks. In 2023, DTP (diphtheria, pertussis and tetanus toxoid-containing vaccine) immunization coverage had yet return to the 2019 levels with 14.5 million infants unable to receive the initial dose (zero-dose children) while about 22.2 million missed the first dose of measles, a departure from the 19.3 million it was in 2019 \cite{vardell2020global,utazi2022assessing}. 

In most developing countries, gender disparity and women's disempowerment are among the major barriers to healthcare utilization including uptake of immunization services \cite{tracey2022does}. Gender inequality causes unequal distribution of resources, power and right both within the household and the society at large and women are often the most vulnerable.
As primary caregivers and decision-makers for their children's well-being, women still face structural barriers in accessing vaccine services as a result of many constraints, including poor or irregular income, gender norms that confine women, and inability to participate in household decision-making or independently decide on healthcare needs \cite{sankar2024gender,tracey2022does}. Limited access to resources or a heavy work burden could restrain their ability to cover indirect costs of vaccination. Women may understand the need to have their children vaccinated and be desirous of doing so, but where they do not participate in household decision-making and the decision-makers do not oblige or refuse permission 
they may lack the ability to follow through with their aspiration. Consequently, social, economic and cultural contexts are essential factors that influence the health outcomes of women and their children \citep{antai2012gender}. 

As a patriarchal society, Nigeria faces gender segregation and stratification that provide material advantages to men while inadvertently constraining women \cite{makama2013}. Cultural and religious beliefs also reinforce social segregation, ensuring that men are prepared for leadership and decision-making, while women are often confined to domestic chores, leading to 
lose of self-esteem in their adult life \cite{aderinto2001,mensah2023}. 
However, women with influence on decision-making within their households are able to provide adequate resources for their children's well-being including getting them immunized \citep{singh2013maternal,woldemicael2010women,amoah2023influence}. 
Thus, modern strategies, policies, and interventions aimed at promoting healthy lives of all children can reasonably be based on gender equity and women's empowerment \citep{dimbuene2018women,seidu2022women,tracey2022does}. 
We postulate that women's empowerment regarding healthcare utilization and participating in household decision-making could impact on adoption of childhood vaccination differently depending on the child's location of residence. Thus, we aim to demystify the space-time patterns of vaccination coverage based on these two dimensions of women's empowerment in Nigeria.

{Spatio-temporal models have been used to highlight 
 inequities in vaccination coverage in Nigeria. For instance, Utazi et al. \citep{Utazi2020Geospatial} 
reported pronounced spatial heterogeneity and persistent low-coverage regions of measles vaccine coverage, Sbarra et al. \citep{sbarra2023estimating} 
estimated DPT coverage in conflict-affected Borno State, demonstrating that geospatially modeled estimates can be valuable in contexts where insecurity limits representative sampling while Utazi et al. \citep{utazi2023mapping} mapped vaccine coverage at high spatial resolution, and linked vaccination coverage to measles incidence. More recently, Aheto et al. \citep{aheto2023geospatial} examined childhood vaccination coverage at district level, before and during the COVID-19 pandemic, and found no substantial changes over time and space during the study period. Our study extends existing research by integrating women's empowerment indicators as key spatial covariates to examine how gender-related factors 
interact with spatial and temporal patterns of childhood vaccination coverage in Nigeria. Specifically, we assess five vaccination outcomes: Bacillus Calmette-Gu\'{e}rin (BCG), zero-dose status, completion of the DPT series, the first dose of the measles-containing vaccine (MCV1), and receipt of all basic vaccinations. This study is the first to integrate a composite Women's Empowerment Index within a spatio-temporal framework to assess how empowerment interacts with local context and time-varying conditions to shape multiple dimensions of childhood immunization coverage in Nigeria. The findings provide new insights into how varying levels of women's empowerment shape disparities in childhood immunization, thereby informing policies aimed at promoting equitable vaccine coverage nationwide.}

\subsection{Measurement of women's empowerment}
There are irregularities in the definition of women's empowerment but it is operationally guided by the principle of civil and social rights, and relates to enhancing capabilities, equal access to resources and opportunities, and having the agency to use these to make strategic choices and decisions \citep{malhotra2005women,kazembe2020women}. Kabeer \cite{kabeer1999resources,kabir2005gender} regarded empowerment as a pathway through which women gain agency, and agency is considered as the capability to make conscious life choices. Thus, operationally, the idea of women's empowerment has been summarized into three main concepts: (i) Participation in decision-making; (ii) Increased access to productive resources; and (iii) Expanded choices of individual women \citep{kazembe2020women, malhotra2005women,robinson2017interventions}. Household surveys, such as the Demographic and Health Survey (DHS) have collected several variables that when combined, can be useful in measuring different aspects of women's empowerment. However, the numerous indicators involved pose some computational challenges in an attempt to pool them in a modelling setting. Several transformative approaches have therefore been proposed to combine these variables while preserving the required information for measuring different dimensions of women's empowerment. The study by Banks et al. \cite{banks2022geography} used a multivariate cluster analysis to identify empowerment patterns based on responses to relevant DHS questions, and classified women into clusters related to their level of empowerment. Female empowerment index calculated based on the proportion of positive outcomes from different domains of women's empowerment, taking values between zero and one to indicate low to high empowerment, was proposed by Retting et al. \citep{rettig2020female}. Kazembe \cite{kazembe2020women} utilized factor analysis to combine responses from several questions related to various aspects of empowerment. The latent scores generated were divided into tertile (three groups) and used to assign the women into one of low, moderately, or highly empowered category, to facilitate interpretation and allowing for the study of the spatial dimensions of empowerment among Namibian women. The Kazembe's approach was similar to the principal component method used by Ewerling et al. \citep{ewerling2017swper} and have been adopted to investigate the spatial dimension of women's empowerment in Nigeria \cite{adeyemi2024analysis}. 

We relied on Kazembe's \cite{kazembe2020women} framework to create empowerment indicators for household decision-making and healthcare utilization among Nigerian women by combining several relevant variables from the Nigeria Demographic and Health Survey (NDHS) data sets. 
{Although there are several other dimensions of women's empowerment, some of which are derived from indicators that measure domestic violence, contraceptive use, literacy, and earnings, all of which could affect vaccination uptake, we settled for the } two aspects of women's empowerment because of the belief that women who participate in household decision-making would have the agency to mobilize resources to meet their children's needs and would therefore face minimal constraints posed by any direct or indirect cost of immunization. They would be able to move freely, 
which is important to carry out activities that are of potential benefit to their children and would not have to wait for their partners before deciding to vaccinate their children \citep{mohanty2018maternal}. 
Women who could freely decide when, where and how to meet their healthcare needs would be able to translate this benefit to their children without hindrance. We settled for the Kazembe's framework because of its simplicity in creating latent scores for women's empowerment that preserve the information in the original variables and its usefulness in assigning the women into mutually exclusive categories that demonstrate their class of empowerment. Moreover, the approach has shown usefulness in quantifying the geographical patterns of different dimensions of women's empowerment in Nigeria \citep{adeyemi2024analysis}. 




\section{Materials and Methods}
\subsection{Data}
We relied on data from four waves of the NDHS conducted in years 2003, 2008, 2013, and 2018. These national household surveys were conducted following multistage sampling techniques with clusters selected from the sampling frames of the 1999 and 2006 Population and Housing Censuses of the Federal Republic of Nigeria. Details of the sampling procedure and sample sizes for each survey have been reported in \citep{national2004nigeria,national2008nigeria,national2013nigeria,national2018nigeria}. Information regarding vaccination coverage was collected by inspecting the vaccination card shown to the interviewer or from the report provided by the mother, in the case the card is unavailable. We considered whether or not the child had received each of the following vaccines: BCG, DPT-complete (receiving three doses of DPT), MCV1, and all-basic vaccination, defined as a child receiving a dose of BCG, three doses of DPT, three doses of oral polio vaccine (excluding the one at birth), and a dose of MCV. We also included zero-dose (proxied by no DPT). Our analysis was restricted to children aged 12-23 months. {After data cleaning and coding, 17,892 children were included in the analysis. Missingness in the data, which was assessed to be low (typically $<2\%$) and missing completely at random, was handled using complete case analysis (list-wise deletion). This approach provides unbiased parameter estimates.} 


\subsection{Empowerment index}
To construct the empowerment index for decision-making and healthcare utilization, we extracted variables similar to those used by Kazembe \cite{kazembe2020women} and Adeyemi and Gayawan \cite{adeyemi2024analysis} from the survey datasets {and utilized Kazembe's \cite{kazembe2020women} method for index construction}. For decision-making, the respondents were asked to state who has the final say on (i) large household purchases (ii) respondent's healthcare (iii) what to do with money husband earns, and (iv) visits to family or relatives. The responses were originally designated as 1: if the woman made the decision; 2: if decided together with her husband/partner; 3: if the husband/partner alone made the decision; and 4: if decided by someone else. We recoded these to 1: if the respondent did not take part in the decision; 2: if she participated in the decision, and 3: if it was a sole decision by the woman. In the case of healthcare utilization, the questions were framed to measure the barriers faced by women when trying to access healthcare needs for themselves. The questions were designed as: getting medical help for yourself: (i) getting permission to go, (ii) getting money needed for treatment, (iii) distance to a health facility (iv) not wanting to go alone. The responses were coded 1: if the woman considers it a big issue, and 2: if she does not consider it a big issue. We perform factor analysis for each set of responses to generate latent factor scores. {After rigorous examination and downstream validation, we selected the first score, which explained much of the variations in the dataset. For interpretation purposes, the score was further divided} into tertiles, producing a three-level ordinal scale that assigns the women according to their class of empowerment. The first tertile corresponds to not empowered, the second, moderately empowered, while the third indicates highly empowered. {The tertile was determined based on the non-linear relationship between the proportion of vaccinated children and intervals of the empowerment scores. The tertiles were computed separately for each survey data before combining. }Table \ref{distr} presents the number of children included from each year of the surveys and the distribution of women in each category of the empowerment indicators as computed. 

\subsection{Data Availability}
The dataset analysed in this study are publicly available upon request from The DHS Program (\url{https://dhsprogram.com/}).
 \subsection{Ethic Statement}
 This study is based on a secondary data obtained from The Demographic and Health Survey (The DHS Program) after approval to use the data was granted by the Program. 
\begin{table}[H]
	\centering
	\caption{Percentage distribution of children in the study and for the mothers based on their class of empowerment}

\begin{tabular}{lcc}
	\hline
	Variable & Number of children/women & Percentage \\
	\hline
	\textbf{Year} &            &            \\
	
	2003 &      1,012 &       5.66 \\
	
	2008 &      5,006 &      27.98 \\
	
	2013 &      5,790 &      32.36 \\
	
	2018 &      6,084 &         34.00 \\
	
	\textbf{Decision-making} &            &            \\
	
	Not empowered &      8,170 &      45.66 \\
	
	Moderately empowered &      4,562 &       25.50 \\
	
	Highly empowered &      5,160 &      28.84 \\
	
	\textbf{Healthcare utilization }&            &            \\
	
	Not empowered &      5,533 &      30.92 \\
	
	Moderately empowered &      5,059 &      28.28 \\
	
	Highly empowered &      7,300 &       40.80 \\
	
	\textbf{Total }&     17,892 &        100 \\
	\hline
\end{tabular}  
\label{distr}
\end{table}

\subsection{Statistical Modelling}
\subsubsection{Vaccine response model}

Let $Y_{ijt} \in \{0,1\}$ denote a random variable for a given vaccine indicator, where $Y_{ijt} =1$ indicates that a child $i\, (i=1,2,...,n_j)$ in local government $j\,(j=1,2,...,774)$ in year $t\, (t=2003,2008,2013, 2018)$ has received a given vaccine, say BCG. Given the nature of the indicator, we assumed that the probability distribution that describes whether a randomly selected child has received vaccination is Bernoulli distributed. That is, $Y_{ijt}\sim Bernoulli (\pi_{ijt})$, where $\pi_{ijt}$ is the probability of receiving a vaccine. We model the probability of covariates using a logit function such that 
\begin{align}
\begin{aligned}
    Y_{ijt}&\sim \texttt{Bernoulli} (\pi_{ijt})\\
    \pi_{ijt}& = \frac{1}{1+\exp(-\eta_{ijt})},
\end{aligned}
\end{align}
where $\eta_{ijt}$ is a linear predictor of the empowerment exogenous variables and the latent spatial field. We considered the following linear predictor

\begin{align}
\begin{aligned}
    \eta_{ijt} &= \alpha_{jt}+\gamma^{(m,d)}_{jt}\delta^{(m,d)}_{ijt}+\gamma^{(h,d)}_{jt}\delta^{(h,d)}_{ijt},\gamma^{(m,hc)}_{jt}\delta^{(m,hc)}_{ijt}+\gamma^{(h,hc)}_{jt}\delta^{(h,hc)}_{ijt},\\
    \delta^{(m,d)}_{ijt} &=\begin{cases}
        1& \text{ if child's }i \text{ mother in LGA}\,j \text{ in time }\,t\text{ is moderately empowered for DM}\\
        0& \text{ otherwise,}
    \end{cases}\\
        \delta^{(h,d)}_{ijt} &=\begin{cases}
            1& \text{ if child's }i \text{ mother in LGA}\,j \text{ in time }\,t\text{ is highly empowered for DM}\\
        0& \text{ otherwise,}
    \end{cases}\\
     \delta^{(m,hc)}_{ijt} &=\begin{cases}
        1& \text{ if child's }i \text{ mother in LGA}\,j \text{ in time }\,t\text{ is moderately empowered for HC}\\
        0& \text{ otherwise,}
    \end{cases}\\
        \delta^{(h,hc)}_{ijt} &=\begin{cases}
            1& \text{ if child's }i \text{ mother in LGA}\,j \text{ in time }\,t\text{ is highly empowered for HC}\\
        0& \text{ otherwise,}
    \end{cases}
\end{aligned}
\end{align}
where DM denotes decision-making and HC denotes healthcare utilization. $\alpha_{jt}$ is the intercept, $\gamma^{(m,d)}_{jt}$ is a latent spatiotemporal variable used to model the spatiotemporal pattern of a given vaccine coverage of children given that the mother/caregiver is ``moderately empowered" in terms of ``decision-making". Similarly,  $\gamma^{(h,d)}_{jt}$ is the corresponding latent spatiotemporal variable given that the mother/caregiver is highly empowered in ``decision-making". $\delta^{(m,d)}_{ijt}$ and $\delta^{(h,d)}_{ijt}$ are observed indicators for whether the mother/caregiver of child $i$ in LGA $j$ in year $t$ is moderately or highly empowered in decision-making respectively. $\gamma^{(m,hc)}_{jt},\gamma^{(h,hc)}_{jt},\delta^{(m,hc)}_{ijt},$ and $\delta^{(h,hc)}_{ijt}$ have a similar definition for empowerment in ``health-care". Thus, the spatiotemporal effect $\gamma^{(m,d)}_{jt}$ can be interpreted as the difference between the spatiotemporal effect in LGA $j$ for children with moderately empowered mothers in decision-making and children with no empowered mothers for a given empowerment condition. Similarly, the spatiotemporal effect $\gamma^{(h,d)}_{jt}$ is the difference between the spatiotemporal effect in LGA $j$ for children with highly empowered mothers and children with no empowered mothers in decision-making. A similar interpretation applies to empowerment in healthcare.

\subsubsection{Model specification for the spatiotemporal latent effects}
This work was considered within the Bayesian paradigm. In a Bayesian estimation procedure, model parameters are treated as random variables, and prior distributions are assigned to these parameters. This approach differs from maximum likelihood estimation as it allows for the estimation of the full distributions of the model parameters, thereby enabling a more comprehensive assessment of uncertainties. In the proposed model, there are five distinct parameters, $\alpha,\gamma_{jt}^{(m,d)},\gamma_{jt}^{(h,d)},\gamma_{jt}^{(m,hc)},\gamma_{jt}^{(h,hc)}$. We adopted the \texttt{R-INLA} package in \texttt{R} \cite{lindgren2015bayesian} to estimate these parameters. 

We specify $\alpha_{jt}\sim N(0, \sigma_\alpha^2),\,\forall j,t$. The hyperparameter $\sigma_\alpha^2 = 1000$ was assumed. Let $\boldsymbol\gamma^{(k)}_t = (\gamma^{(k)}_{1t},\gamma^{(k)}_{1t},...,\gamma^{(k)}_{774t})',k\in \{(h,d),(m,d),(h,hc),(m,hc)\}$, 
we assumed an intrinsic conditional autoregressive model (ICAR) for $\gamma^{(k)}_{jt},k\in\{h,m\}$ \cite{besag1991bayesian}. That is, the conditional prior distribution is given as  
\begin{align}
\begin{aligned}
   \gamma_{jt}^{(k)}\mid \boldsymbol \gamma_{-jt}^{(k)}\sim N\Big(\frac{\sum_{j\in \mathcal N_j}\gamma_{jt}^{(k)}}{\#\mathcal N_j},\frac{1}{\tau_\gamma^{(k)}\# \mathcal N_j}\Big),
\end{aligned}
\end{align}
where $\mathcal N_j$ is the index set of LGA(s) sharing neighbors with LGA $j$, and $\#\mathcal N_j$ is the total number of neighbors of LGA $j$. $\boldsymbol \gamma^{(k)}_{-jt}$ is vector $\boldsymbol\gamma^{(k)}_t$ but excluding the component $\gamma^{(k)}_{jt}$ of the vector. The precision parameters $\log(\tau_\gamma^{(k)})$ is assigned a log gamma distribution; that is, $\log(\tau_\gamma^{(k)})\sim \log\text{gamma}(1,5e^{-4})$. 
In addition, we assumed an autoregressive model of order one (AR1) for the temporal component of $\boldsymbol \gamma^{(k)}_{t}$. That is, we assumed that 
\begin{align}
    \boldsymbol \gamma^{(k)}_{t} = \rho\boldsymbol \gamma^{(k)}_{t-1}+
    \boldsymbol\epsilon_t^{(k)},\,\,\, \boldsymbol\epsilon_t^{(k)}\sim\mathcal N(\mathbf 0,\tau^{-1}\mathbf I),
\end{align}
where $\boldsymbol\epsilon_t^{(k)}$ is a vector of independent error terms, $\boldsymbol\tau$ is the precision for the error terms, and $\rho (\mid\rho\mid<1)$ is the temporal autocorrelation coefficient. In \texttt{R-INLA}, the internal representation of the hyper-parameters are as follows $\theta_2=\log\left(\frac{1+\rho}{1-\rho}\right)$, $\kappa = \tau(1-\rho^2)$, and $\theta_1=\log\kappa$. Moreover, a Gaussian prior is assigned to $\theta_2 \sim\mathcal N(0,7^2)$ and a log gamma prior is assigned to $\theta_1\sim Loggamma(1,5e^{-5})$.

\subsection{Validation}\label{val}
We validated the proposed model by computing the empirical prevalence and the predicted prevalence per state for each vaccine indicator. For every state $jt$ we predict the posterior probability of a given response event (zero dose, DPT complete, MCV-1, BCG, or All basics) $\pi_{jt}$ for all the local governments within that state and then compute the posterior average per state.  That is, suppose $\mathcal S_l,l=1,2,...,37$ is an index set of all the LGA within state $l$; thus we compute the posterior probability of a child receiving a specific vaccine for state $l$ as 
\begin{align}
   \hat p_{lt} = \frac{1}{\#S_l}\sum_{j\in S_l}^{}\hat \pi_{jt},
\end{align}
where $\hat \pi_{ij}$ is the posterior predicted prevalence obtained from the model and $\#S_l$ is the total number of local governments in that state. Similarly, we computed the empirical prevalence $E_{lt}$ as 
\begin{align}
   \hat E_{lt}= \frac{1}{n_{jt}\times \#\mathcal S_l} \sum_{j\in \mathcal S_l}^{}\sum_{i=1}^{n_j'} y_{ijt},
\end{align}
where $y_{ijt}$ is the vaccine indicator under consideration, $n'_j$ is the total number of children within a state $j$ where data were observed. The proposed model is adequate for describing vaccine coverage in Nigeria if the empirical prevalence estimate ($ \hat E_{lt}$) and the predicted estimate ($\hat p_{lt}$) are close. The closer they are the more appropriate is the proposed model. 


\section{Results}
\subsection{Model Results}
\begin{figure}[H]
		\centering
		\includegraphics[scale=0.8]{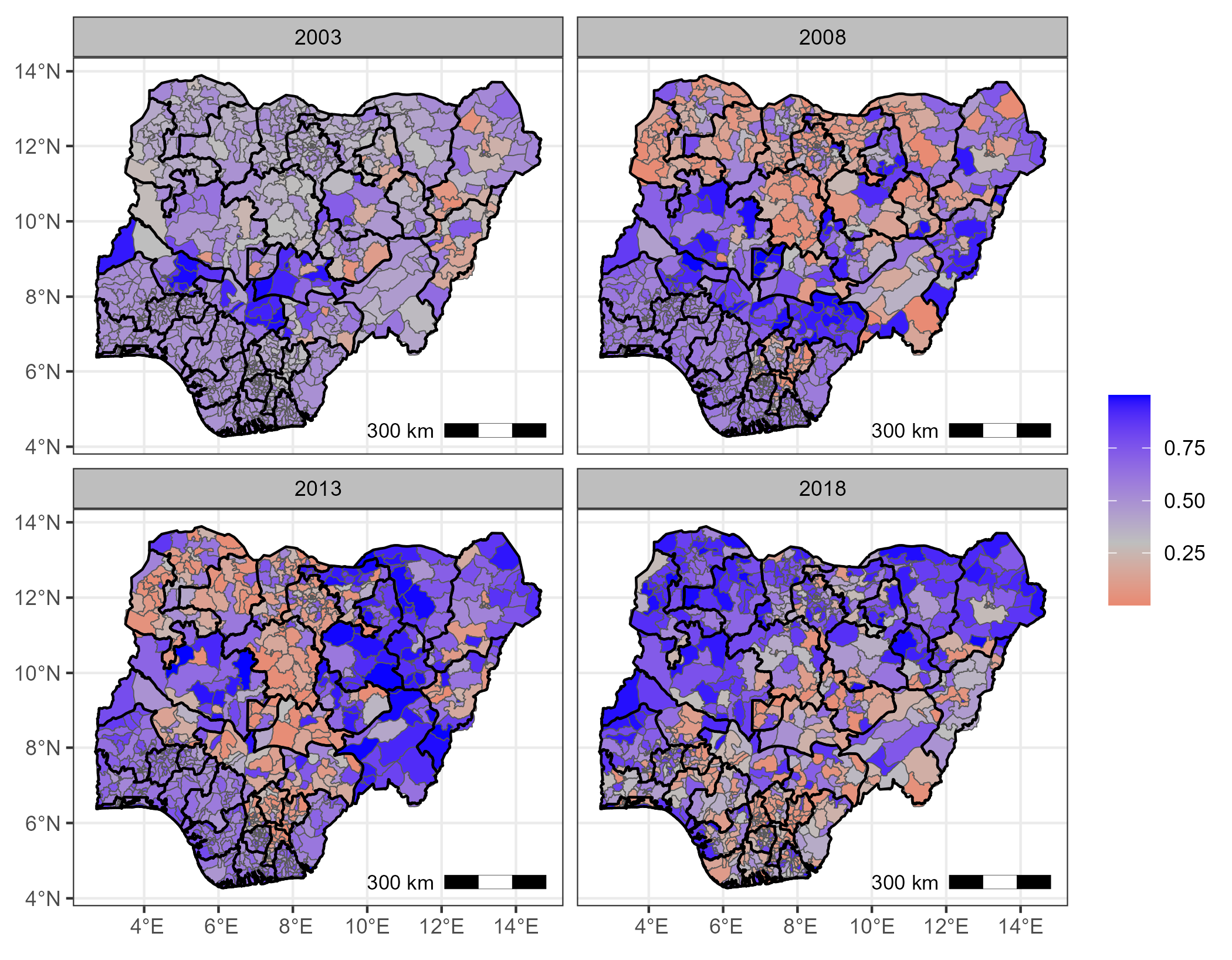}
		\caption{\small Predicted coverage probability $\hat \pi$ of zero dose (no DPT vaccine). Data refer to children aged 12-23 months.}
		\label{pzero}
	\end{figure}

\begin{figure}[H]
		\centering
		{\includegraphics[scale=0.8]{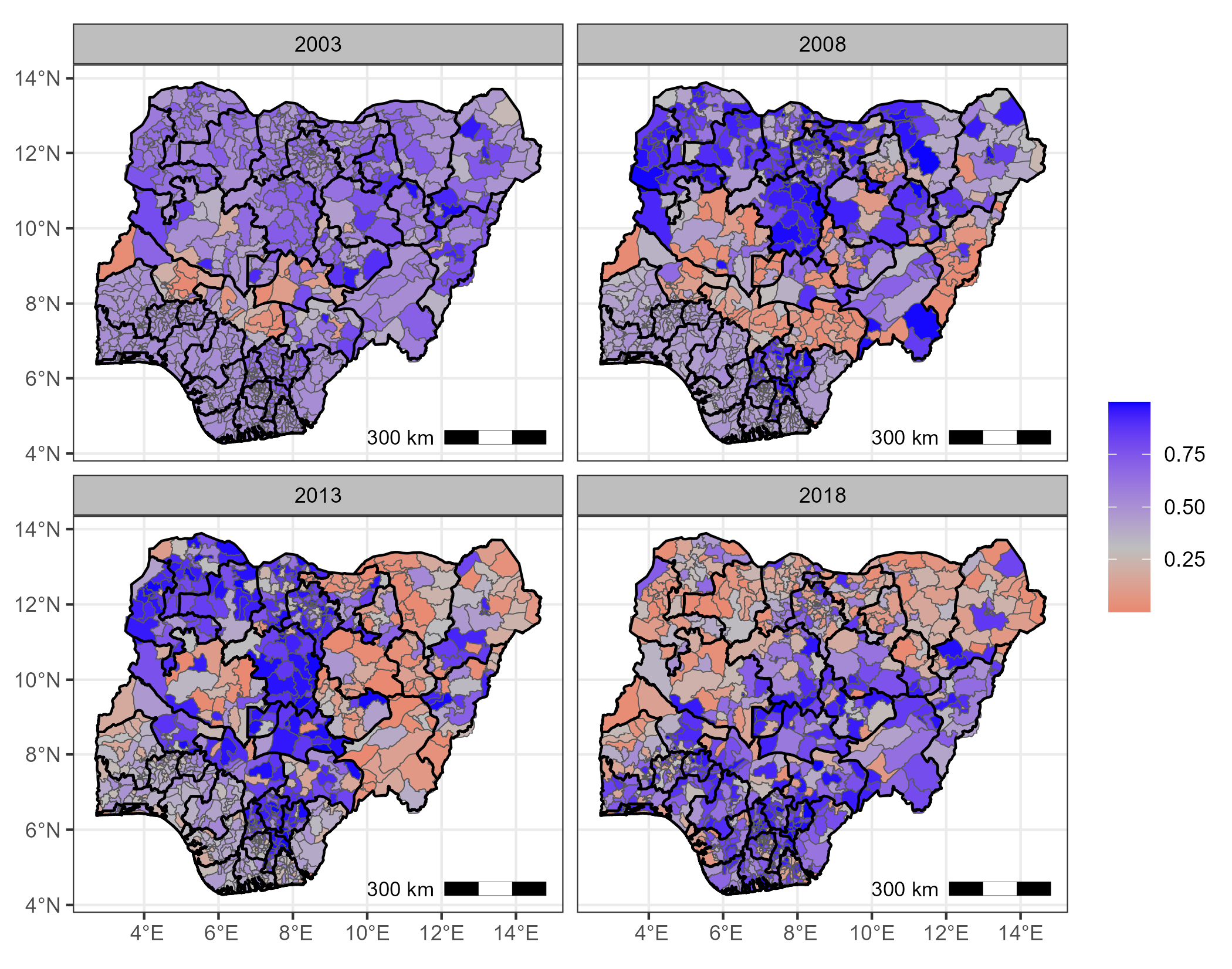}}
		\caption{\small Predicted coverage probability $\hat \pi$ of BCG. Data refer to children aged 12-23 months.}
		\label{pbcg}
	\end{figure}

\begin{figure}[H]
		\centering
		{\includegraphics[scale=0.8]{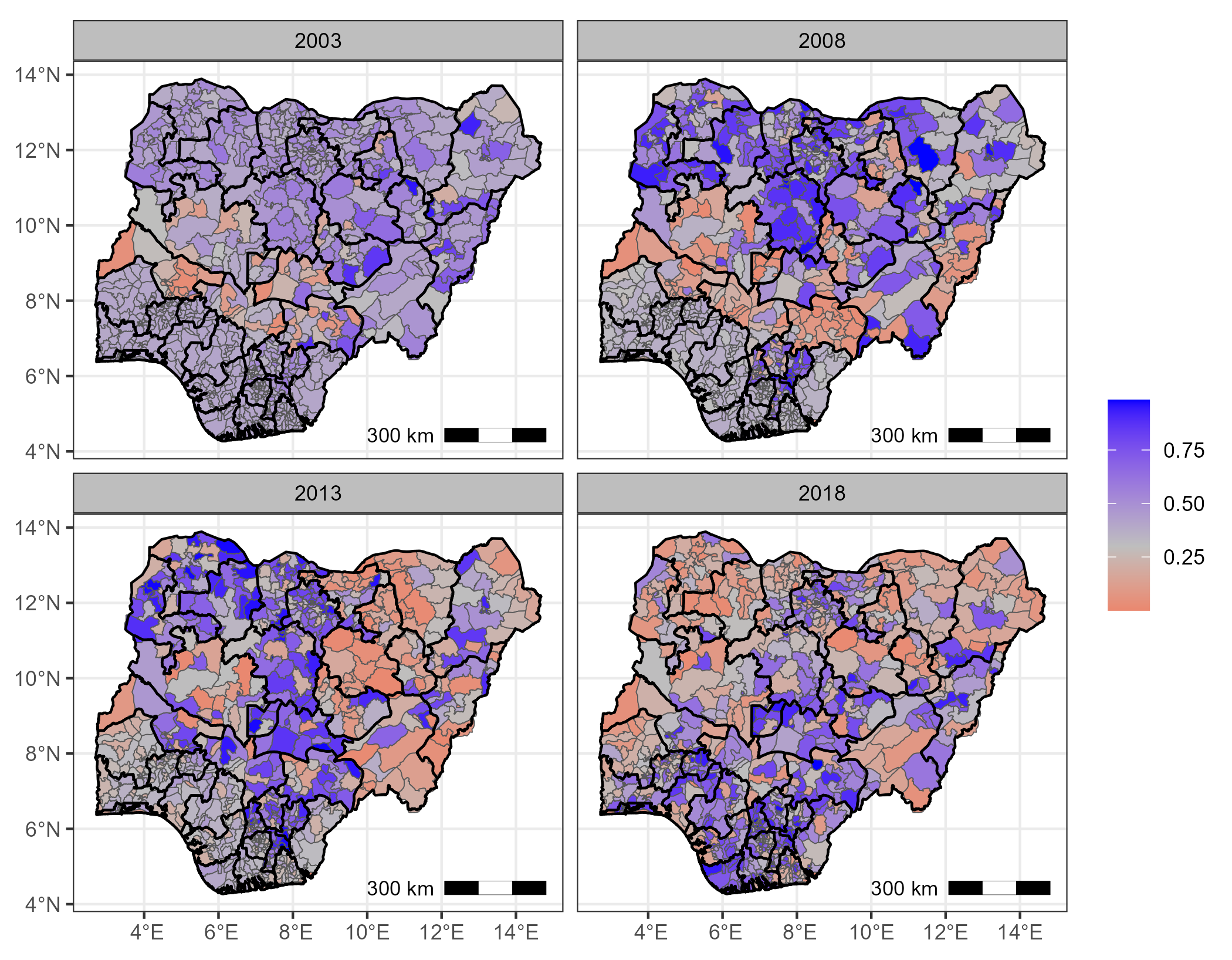}}
		\caption{\small Predicted coverage probability $\hat \pi$ of MCV-1. Data refer to children aged 12-23 months.}
		\label{pmcv}
	\end{figure}

\begin{figure}[H]
		\centering
		{\includegraphics[scale=0.8]{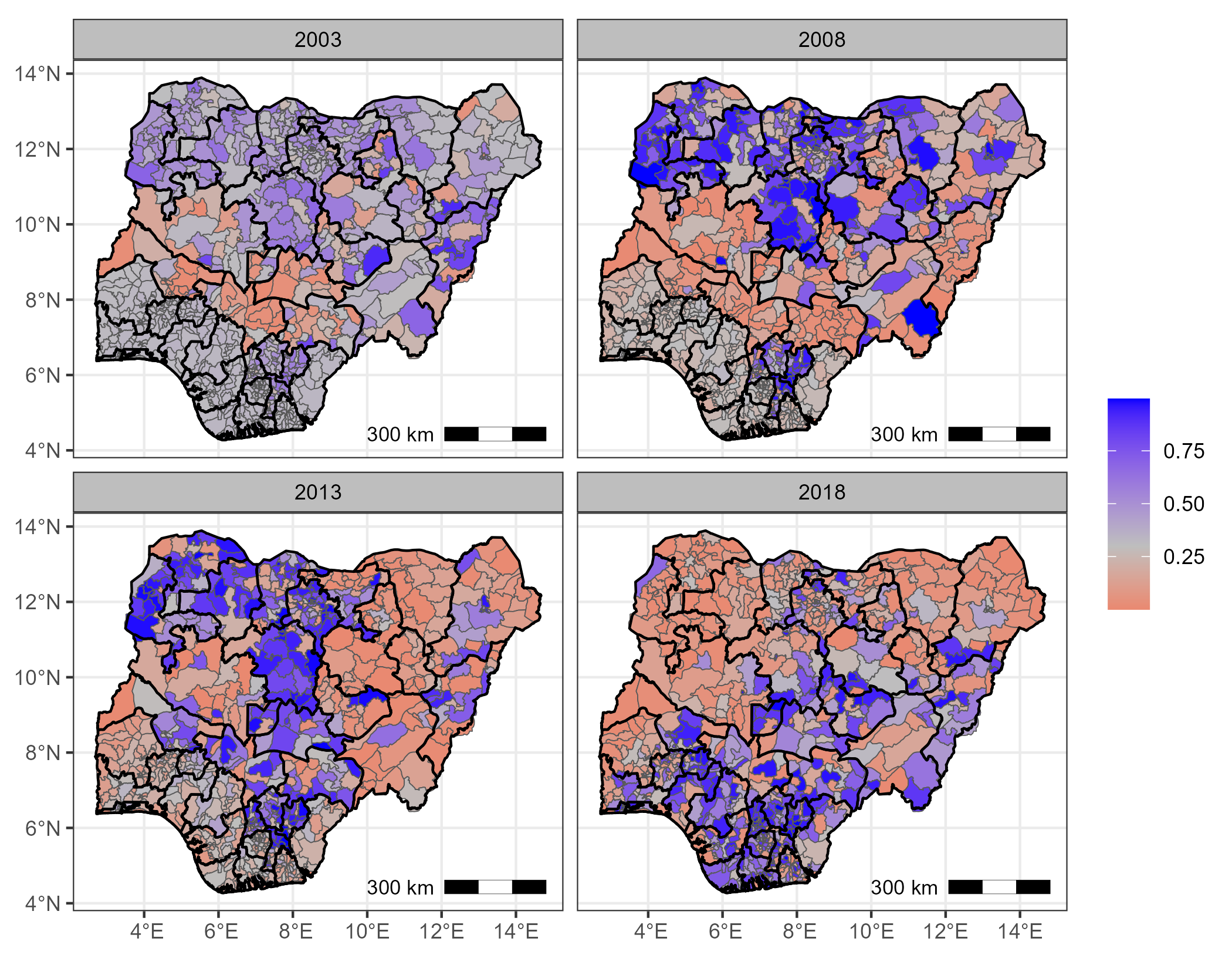}}
		\caption{\small Predicted coverage probability $\hat \pi$ of DPT-complete. Data refer to children aged 12-23 months.}
		\label{dtp}
	\end{figure}

\begin{figure}[H]
		\centering
		{\includegraphics[scale=0.8]{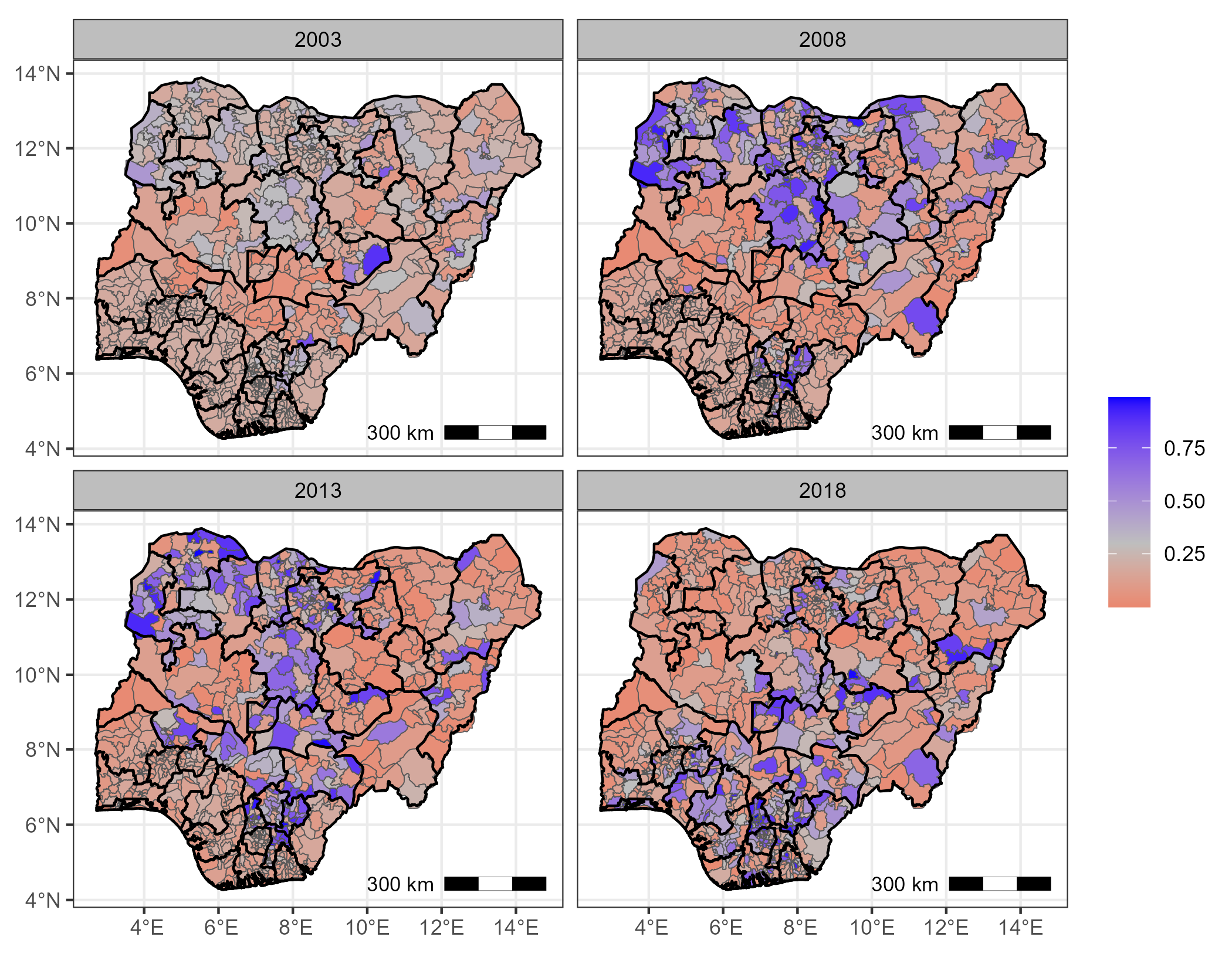}}
		\caption{\small Predicted coverage probability $\hat \pi$ of all basic vaccines (defined as a child receiving a dose of BCG, three doses of DPT, three doses of oral polio vaccine (excluding the one at birth), and a dose of MCV). Data refer to children aged 12-23 months.}
		\label{pall}
	\end{figure}

Fig \ref{pzero} to \ref{pall} present the predicted prevalence of vaccine coverage for zero-dose, BCG, MCV, DPT-complete, and all basic vaccines respectively based on the LGAs of Nigeria. The findings show increasing trends of zero-dose prevalence as prevalence becomes higher in many locations in 2018 when compared with the estimates for 2003 and 2008. For instance, the estimated prevalence was above 75\% in only a few LGAs in Kwara, Kogi, and Nasarawa states, as reflected by the 2003 map but by 2013 and in particular 2018, this had spread to many places, especially in the northern part of the country. But in most southern part, particularly in the south-east region, the estimates are below 25\%, reflecting a dichotomy across space. The estimates for BCG coverage reveal some inconsistency in space-time patterns over the years with coverage being around average in most part of the country in 2003 except for Kwara, Kogi and Nasarawa states where it was lower. In 2008, BCG coverage appeared high in most LGAs in the northern part of the country, lower than 25\% in many LGAs in the central part of the country and about average in most places in the south. In 2013, most LGAs in states like Taraba, Bauchi, Gombe, Yobe, Borno and Niger have lower coverage, while by 2018, a north-south divide was evident with lower prevalence reported in many places in the north. 



The estimates for MCV-1 (Figure \ref{pmcv}) reveal that almost all the LGAs had below 50\% coverage in 2003 but with slight improvement in 2008, where a number of LGAs in the north have higher estimates. This pattern changed in 2013 and in particular in 2018, where a somewhat north-south divide was evident, with prevalence below 25\% in most LGAs in the north, although some locations in Oyo and Ogun states in the south also had lower prevalence comparable to most places in the north. The estimates for DPT-complete (Figure \ref{dtp}) are similar to those of MCV-1 though coverage is much lower, particularly in 2018 in many more places across the country. The maps for all basic vaccinations presented in Figure \ref{pall} show that throughout the period under consideration, coverage for all basic vaccination is low, just about 10\% in almost everywhere in the country, except in a pocket of LGAs.   



\begin{figure}[H]
		\centering
        {\includegraphics[scale=0.455]{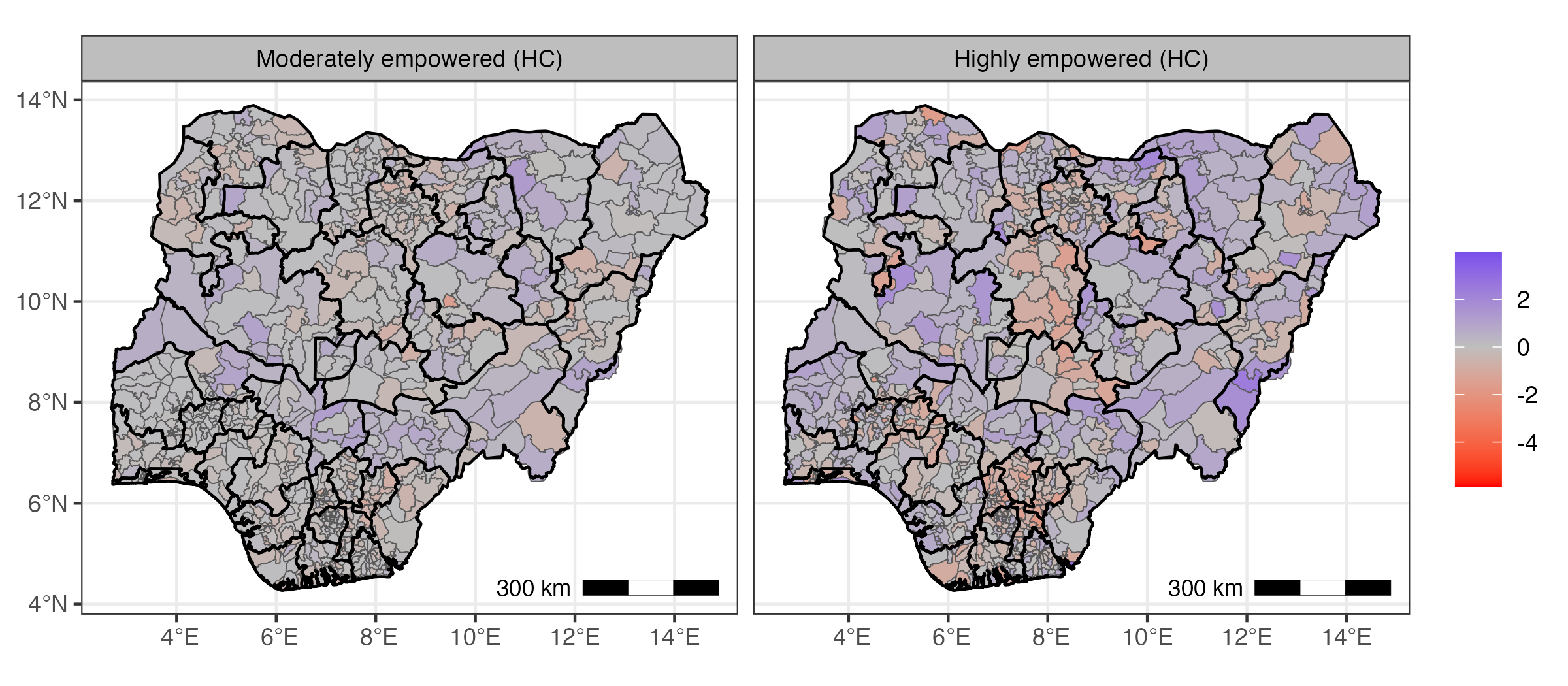}}{\includegraphics[scale=0.455]{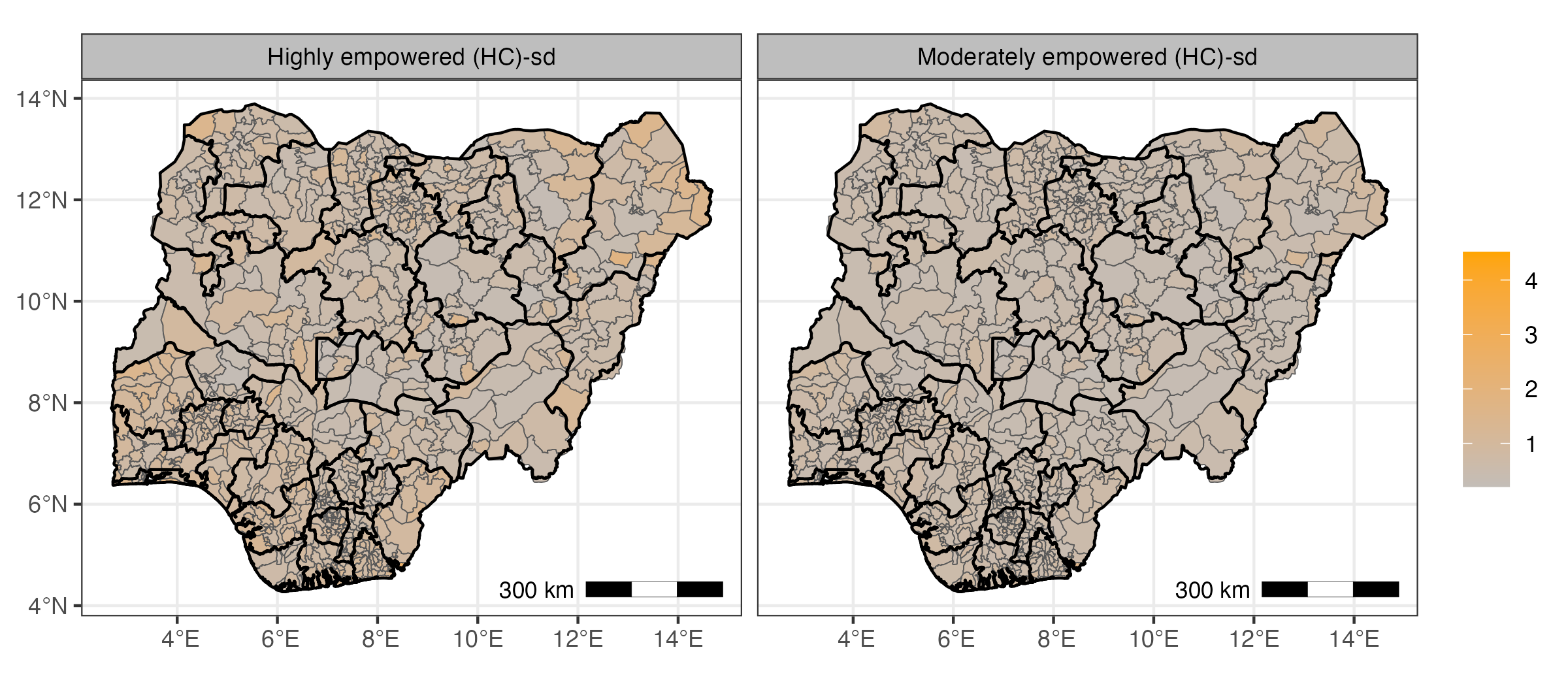}}\\
		{  \includegraphics[scale=0.435]{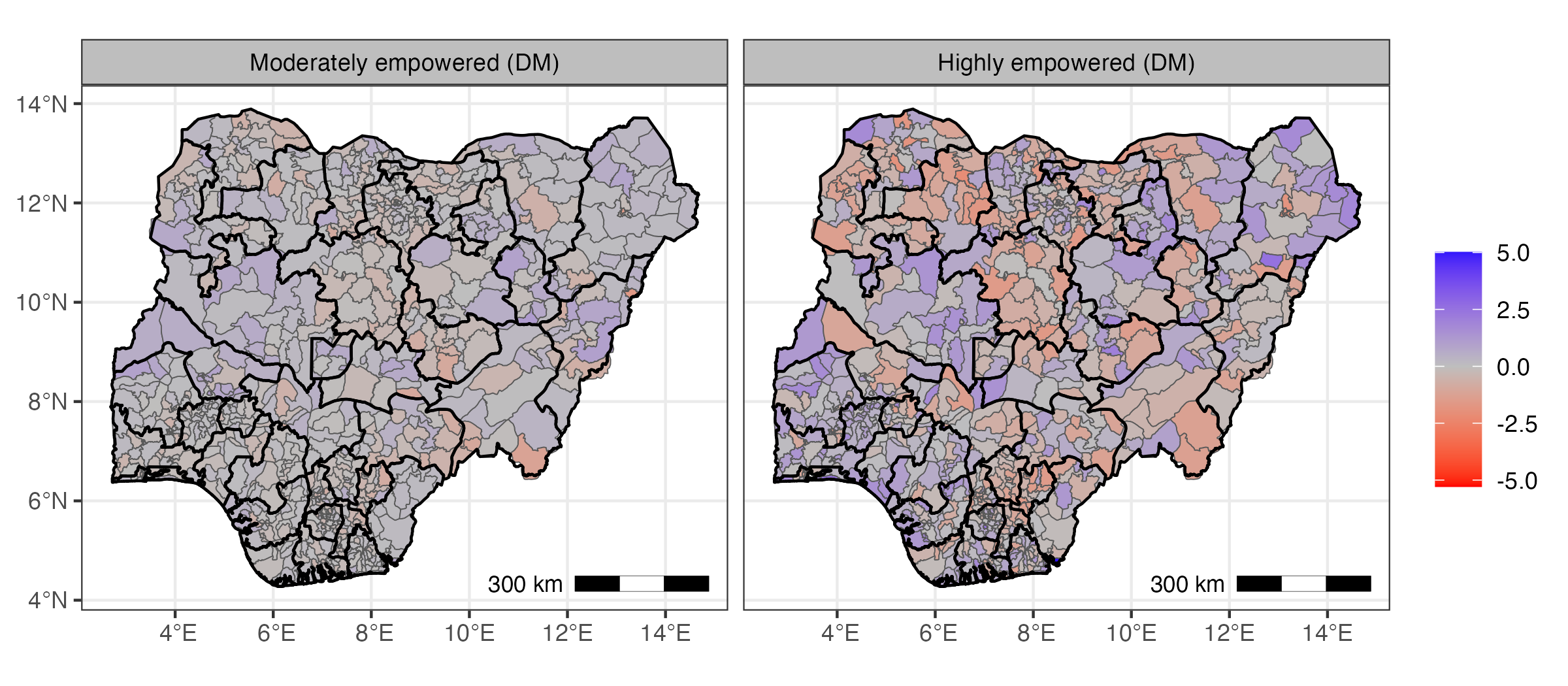}}
        	{  \includegraphics[scale=0.435]{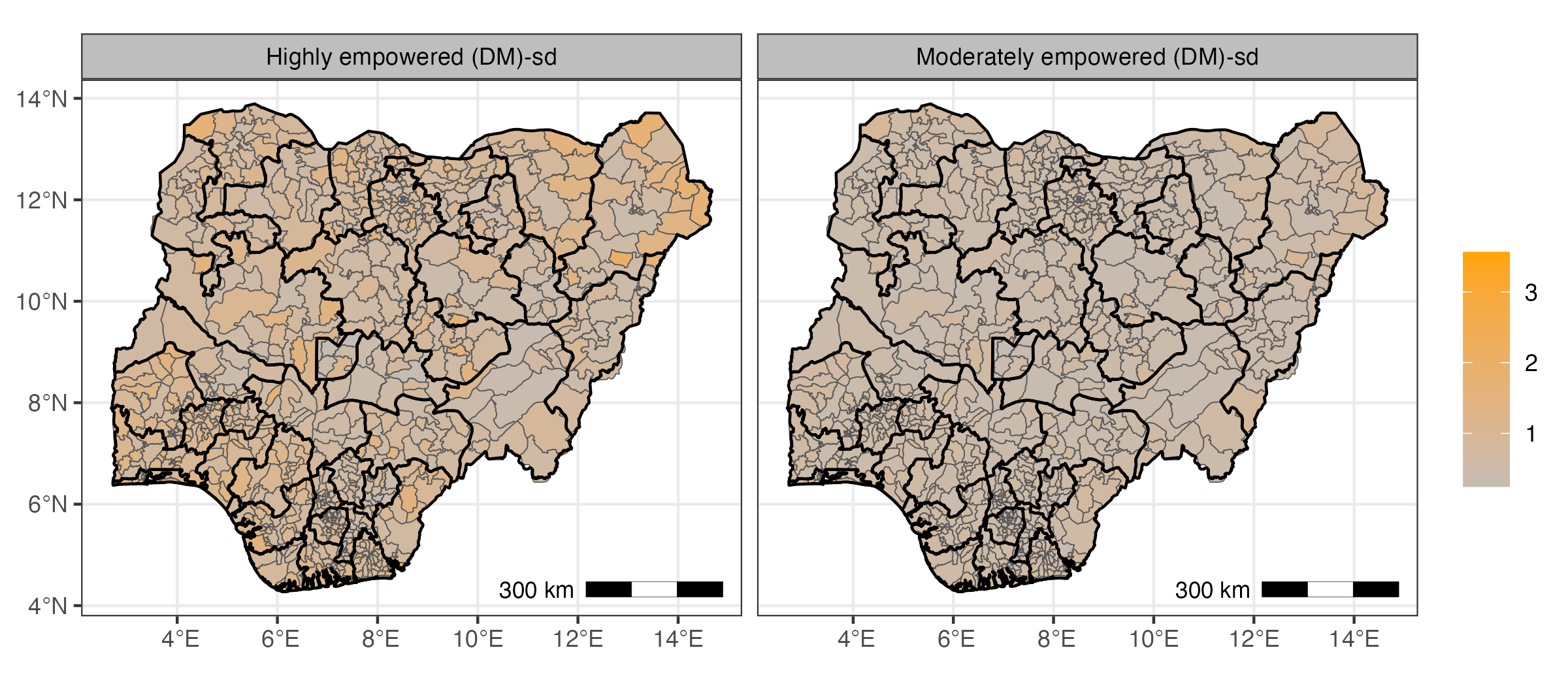}}
		\caption{\small Posterior mean and standard deviation of the effects $\hat \gamma$ of women's empowerment indicators on zero-dose children (no DPT vaccine) coverage. Data refer to children aged 12-23 months.}
		\label{ezero}
\end{figure}
\begin{figure}[H]
		\centering
	{\includegraphics[scale=0.435]{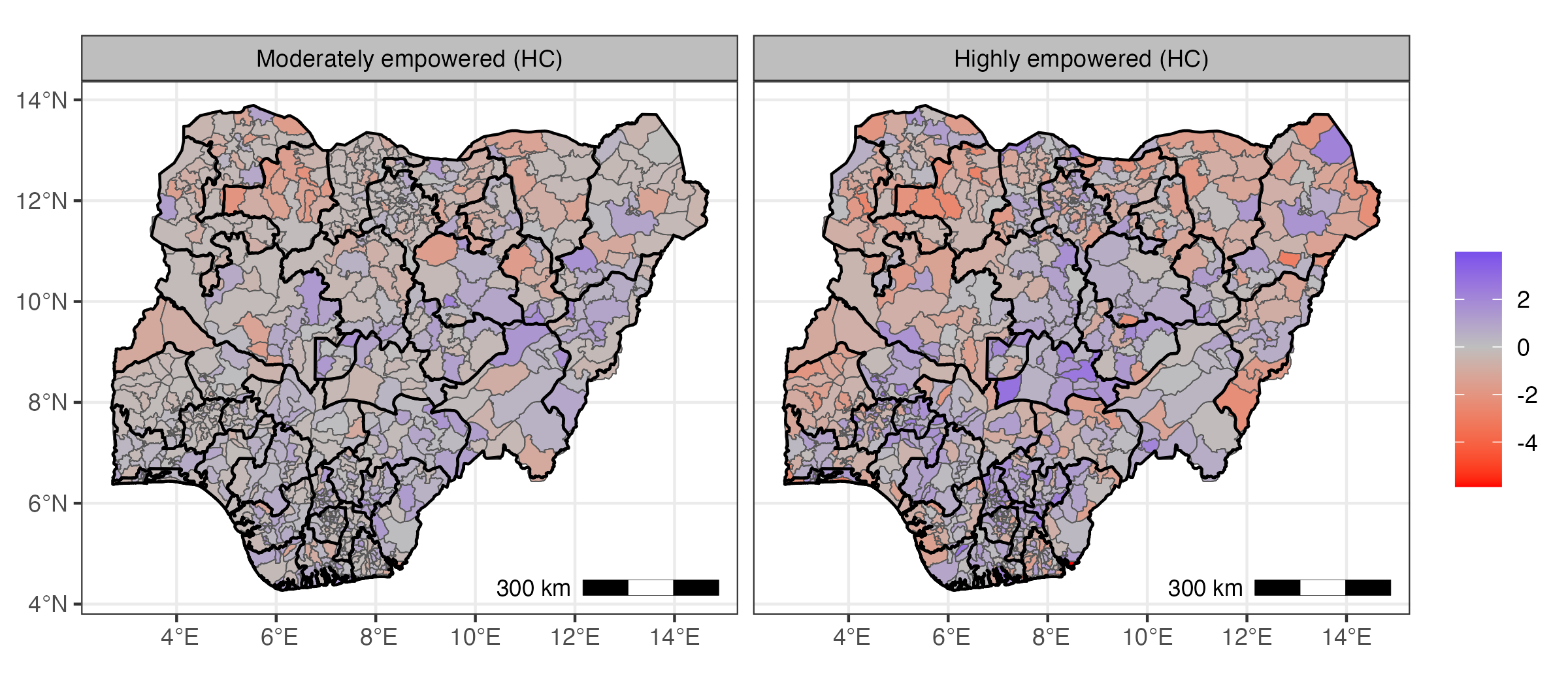}}
    {\includegraphics[scale=0.435]{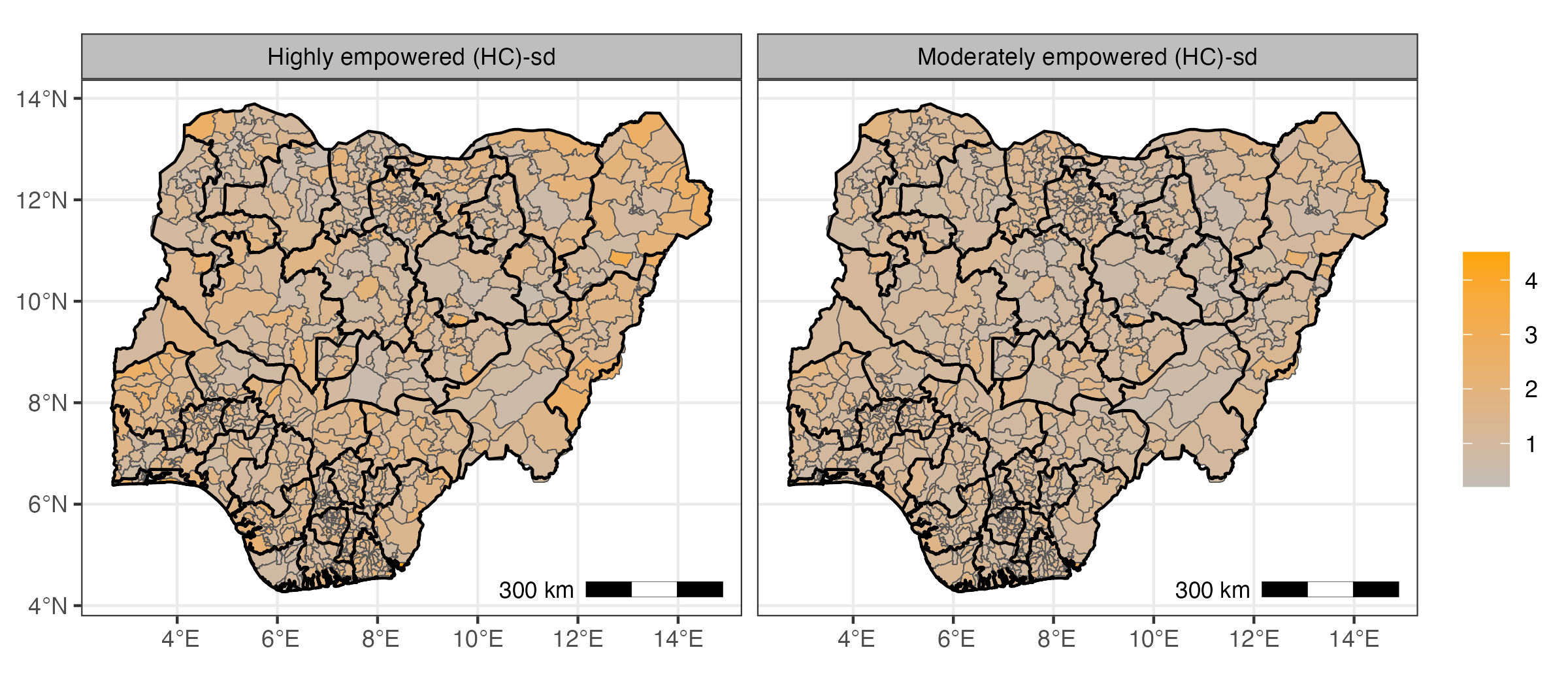}}\\
		{  \includegraphics[scale=0.435]{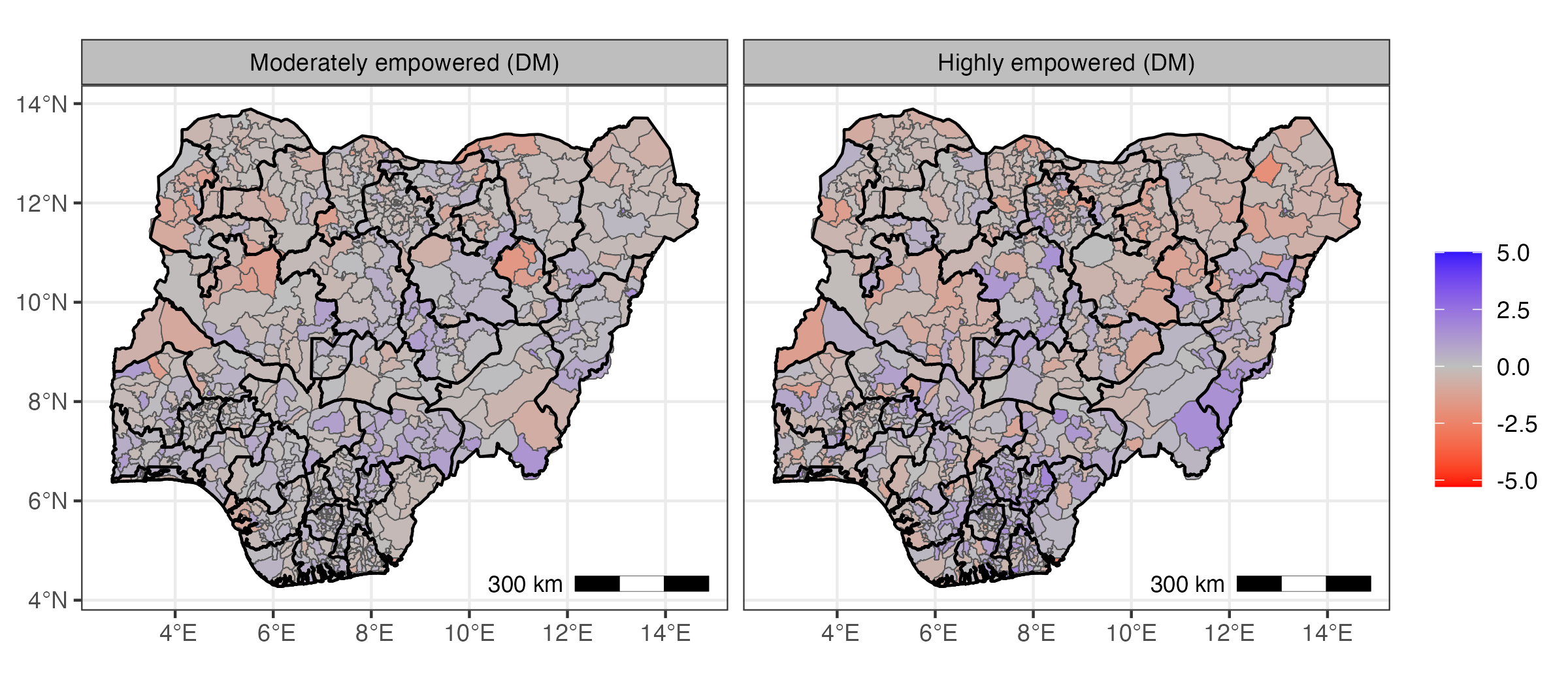}}
        {  \includegraphics[scale=0.435]{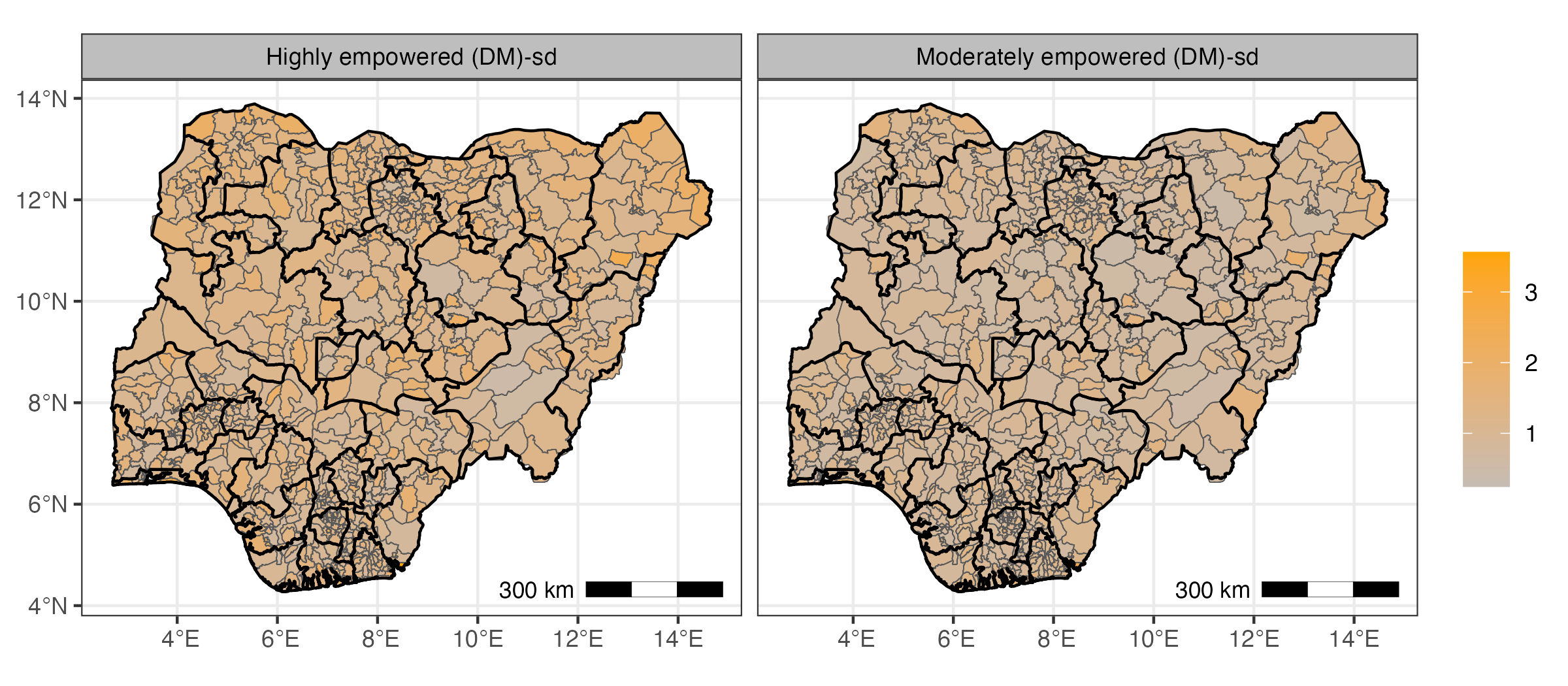}}
		\caption{\small Posterior mean and standard deviation of the effects $\hat \gamma$ of women's empowerment indicators on BCG coverage. Data refer to children aged 12-23 months.}
		\label{ebcg}
 \end{figure}
\begin{figure}[H]
		\centering
        {\includegraphics[scale=0.435]{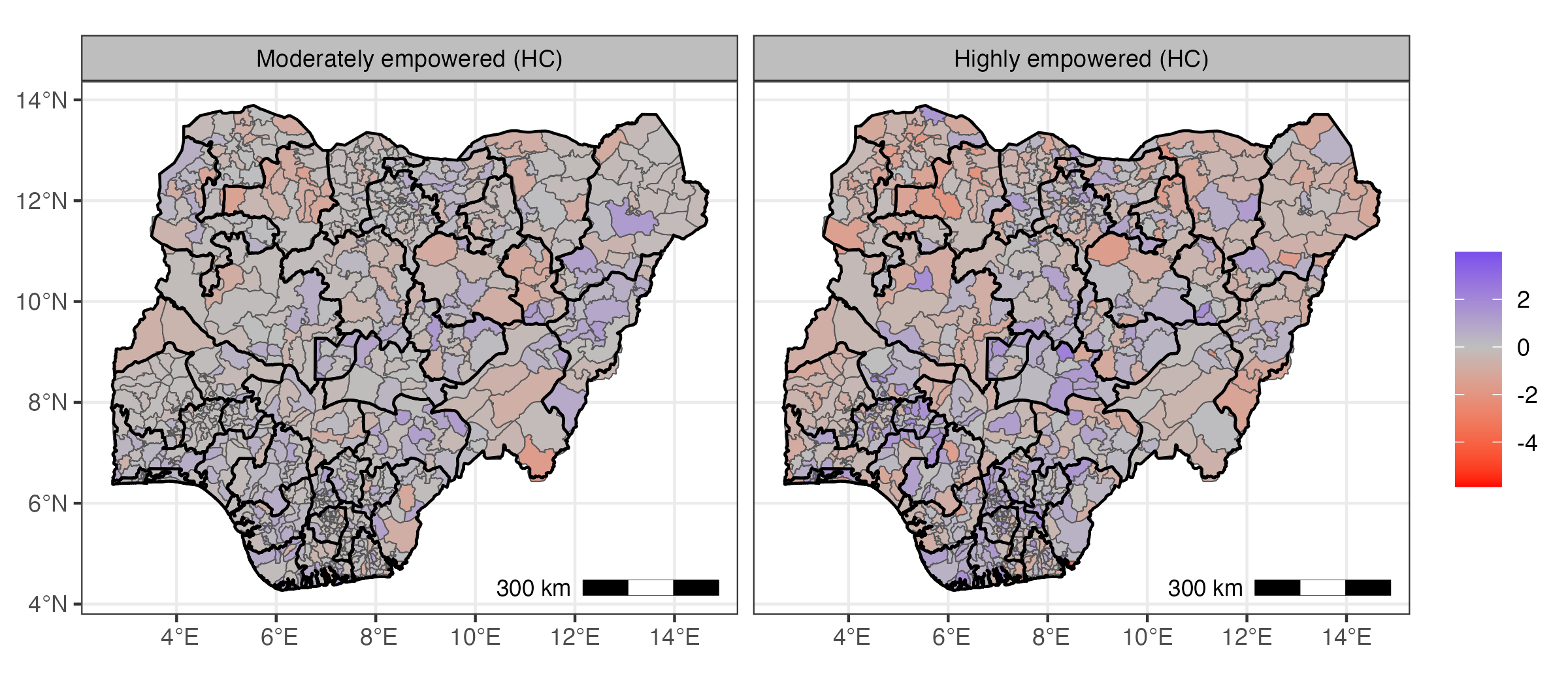}}
         {\includegraphics[scale=0.435]{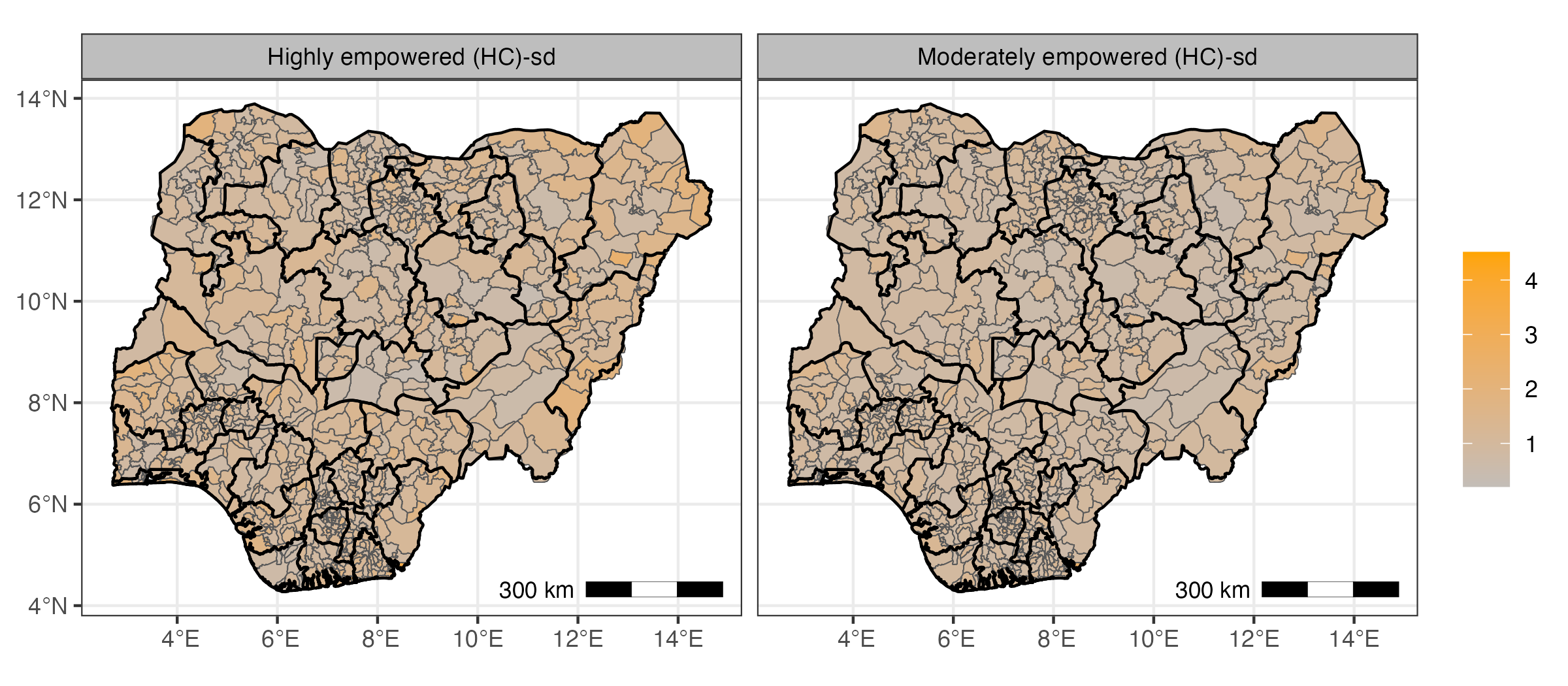}}\\
		{  \includegraphics[scale=0.435]{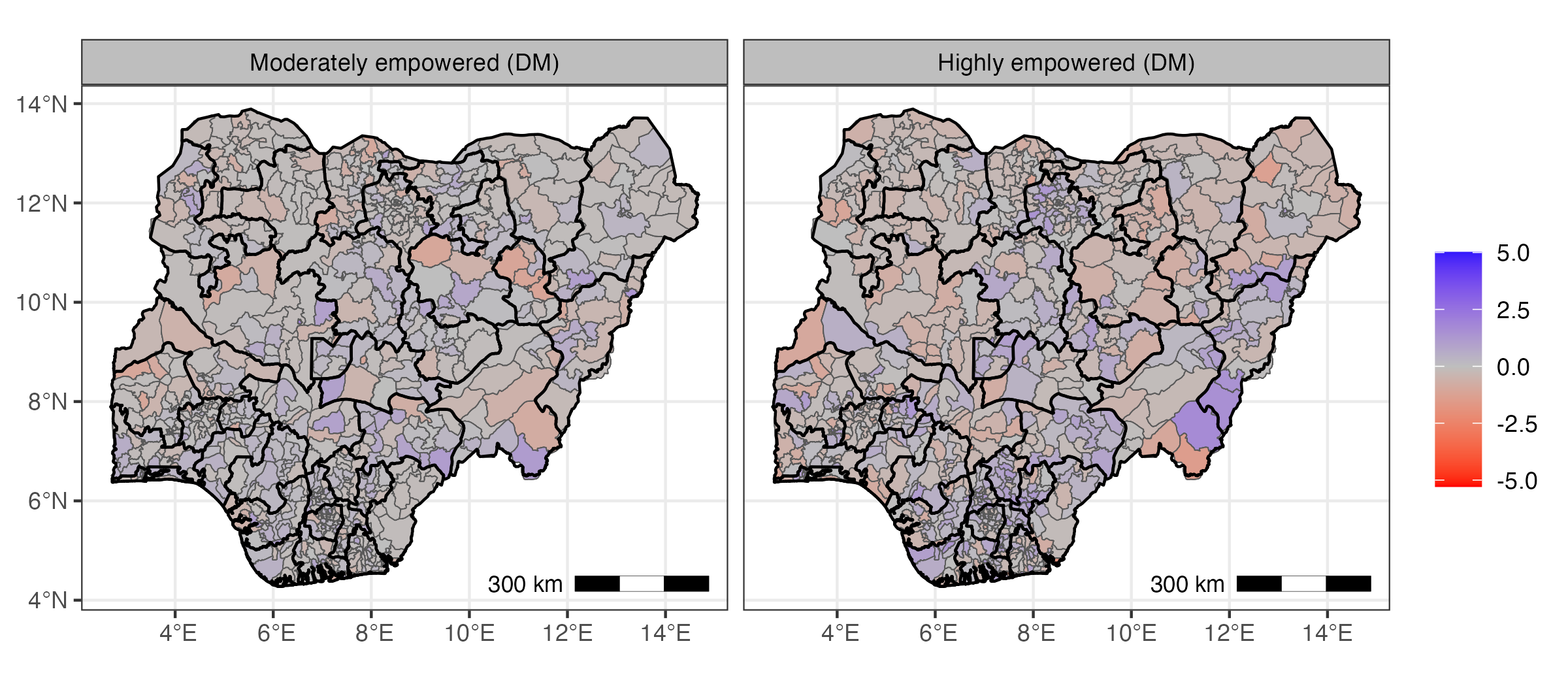}}
        	{  \includegraphics[scale=0.435]{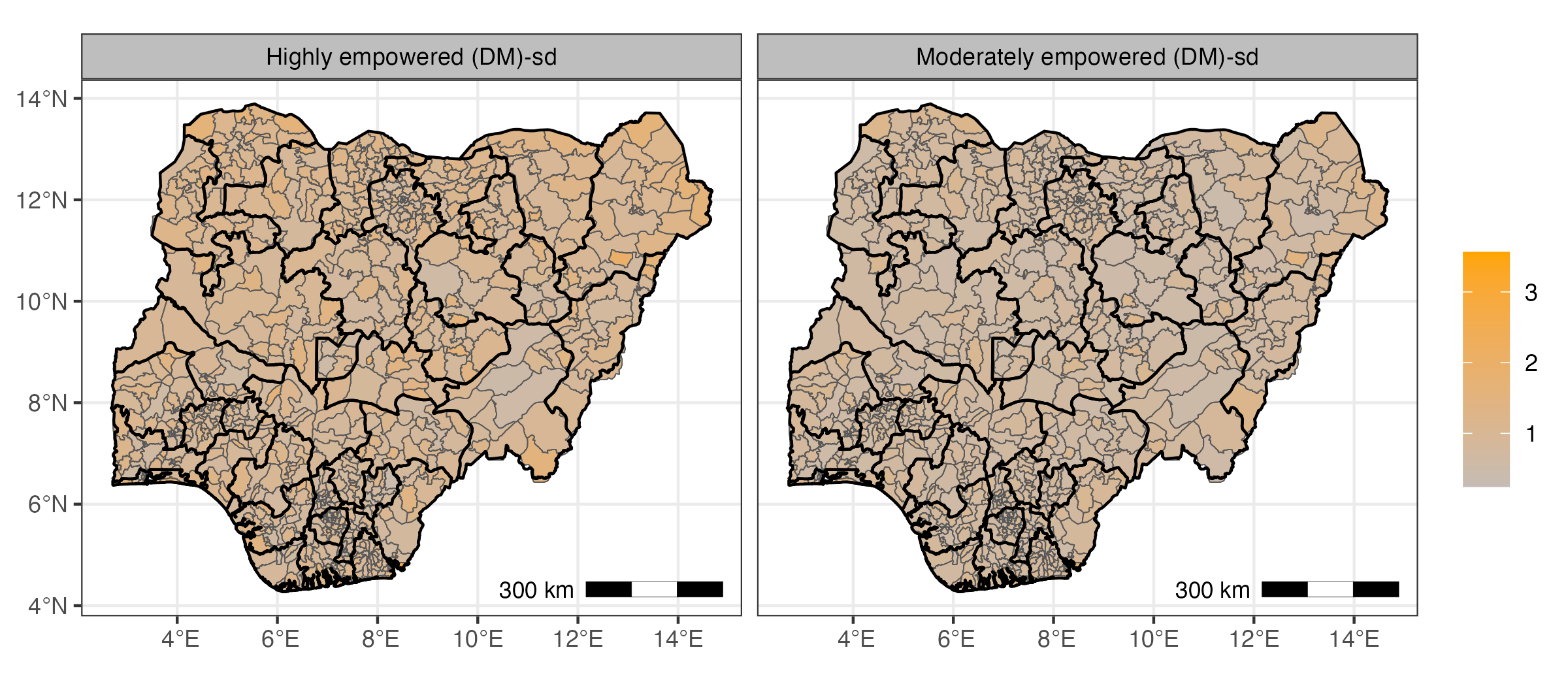}}
		\caption{\small Posterior mean and standard deviation of the effects $\hat \gamma$ of women's empowerment indicators on MCV-1 coverage. Data refer to children aged 12-23 months. }
		\label{emcv}
	\end{figure}
\begin{figure}[H]
		\centering
        {\includegraphics[scale=0.435]{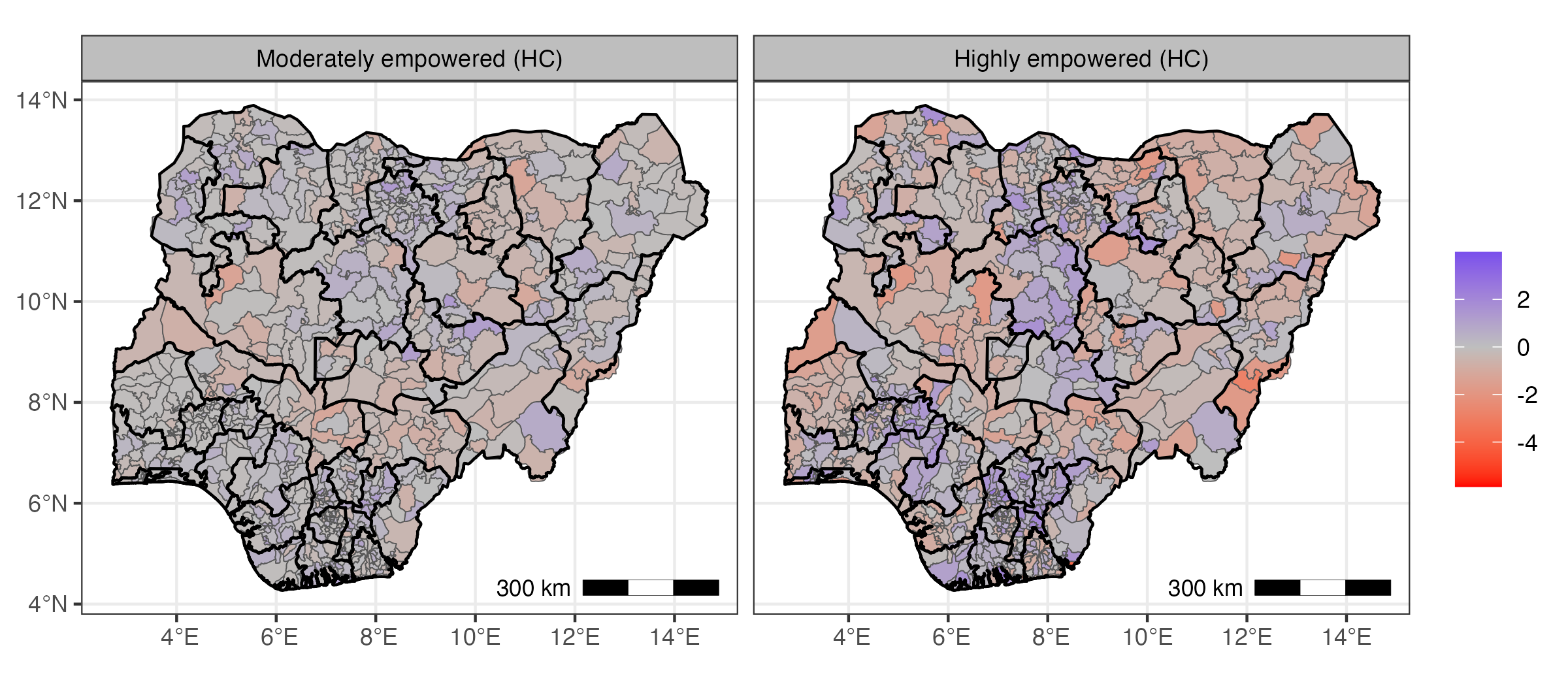}}
         {\includegraphics[scale=0.435]{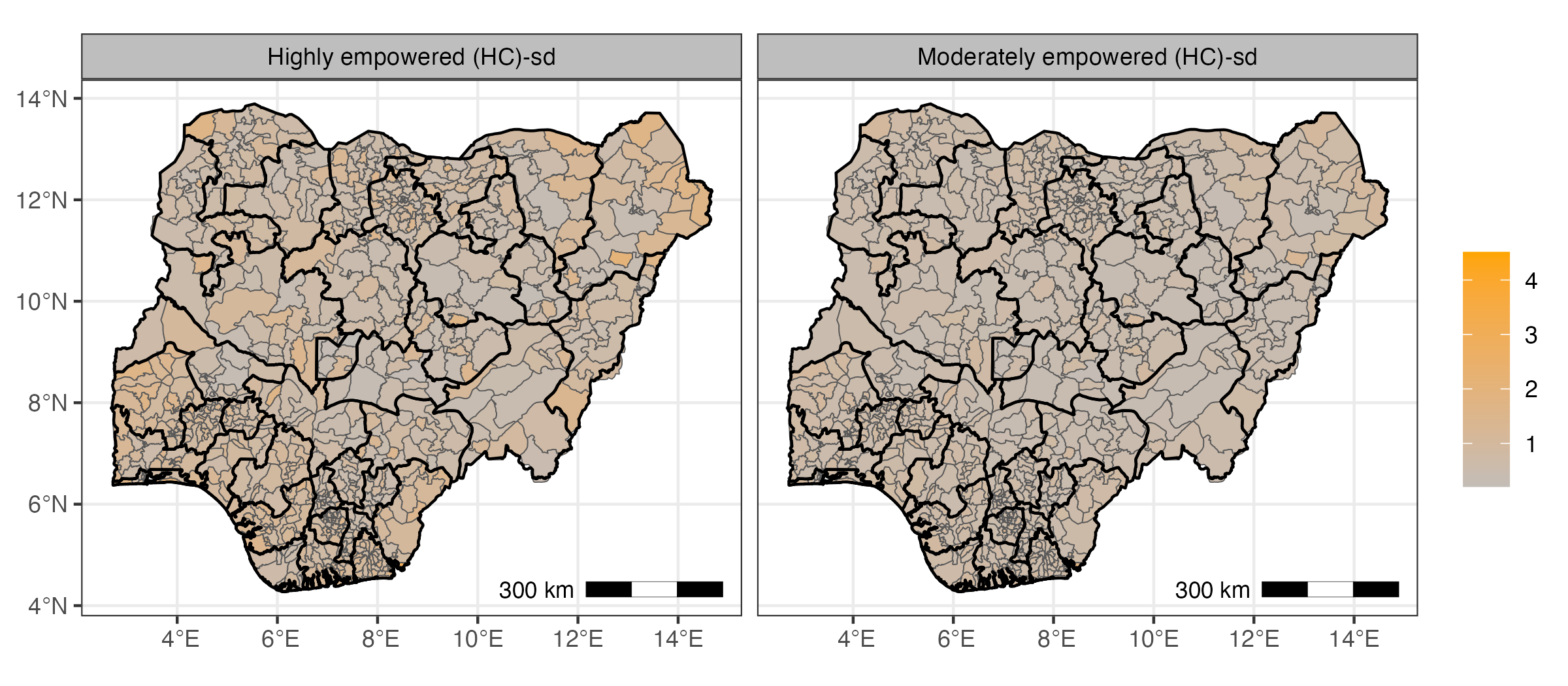}}\\
		{  \includegraphics[scale=0.435]{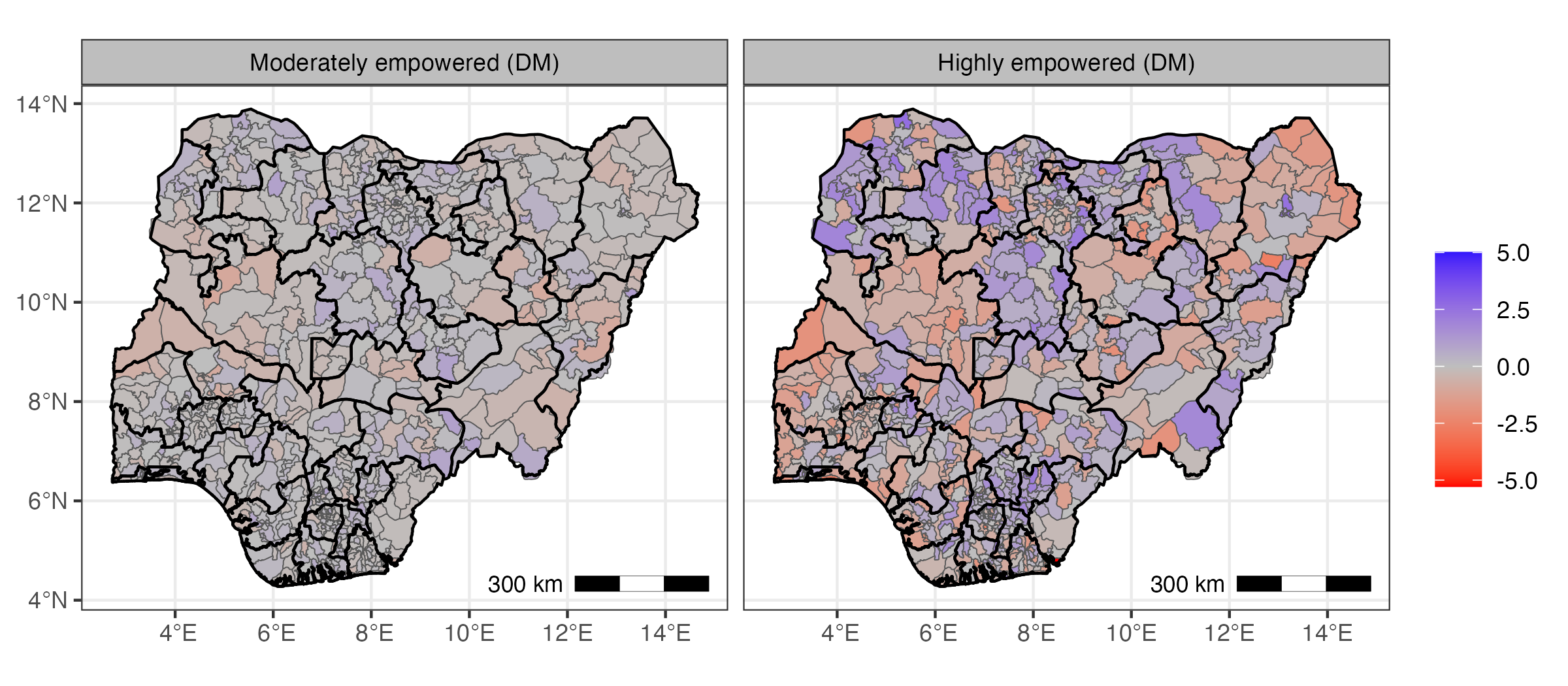}}
        {  \includegraphics[scale=0.435]{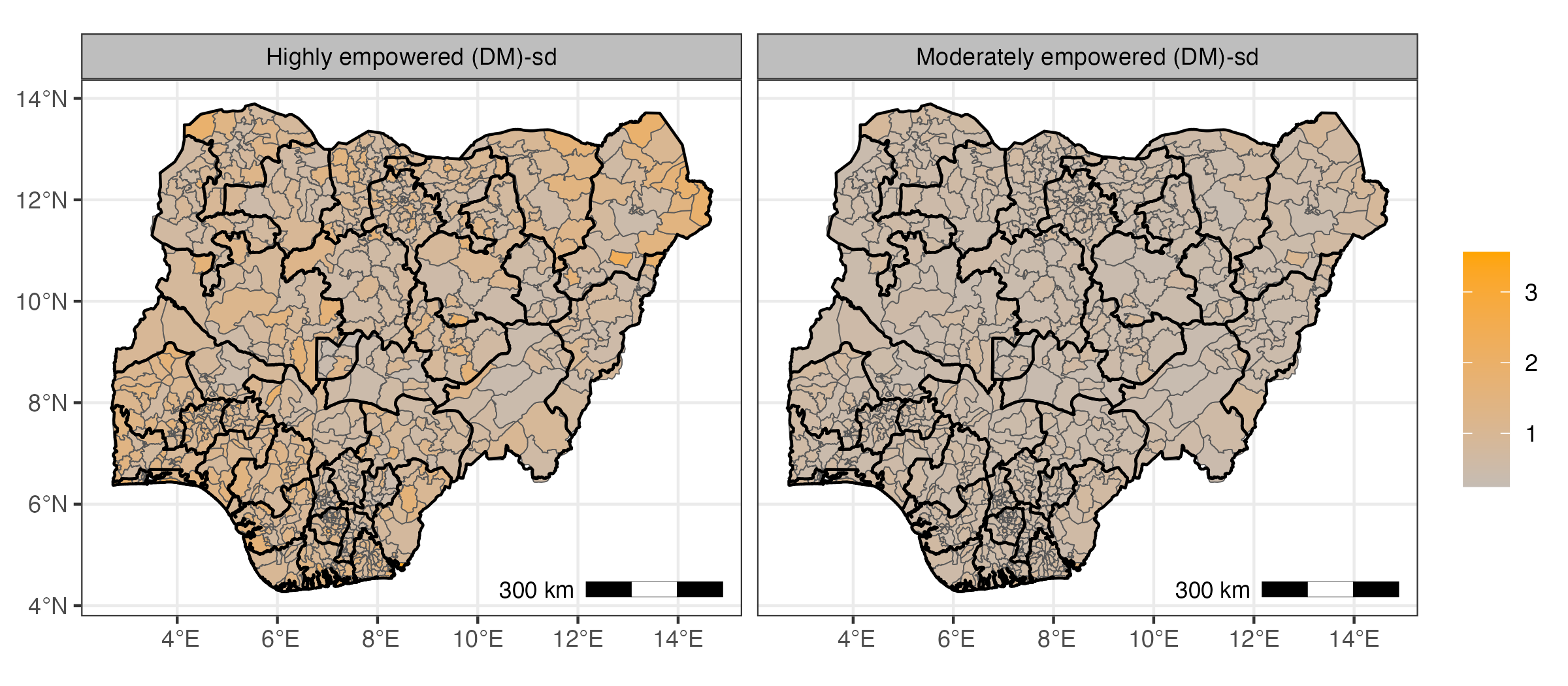}}
		 \caption{\small Posterior mean and standard deviation of the effects $\hat \gamma$ of women's empowerment indicators on DPT-complete coverage. Data refer to children aged 12-23 months.  }
		\label{edpt}
	\end{figure}
\begin{figure}[H]
		\centering
        {\includegraphics[scale=0.4]{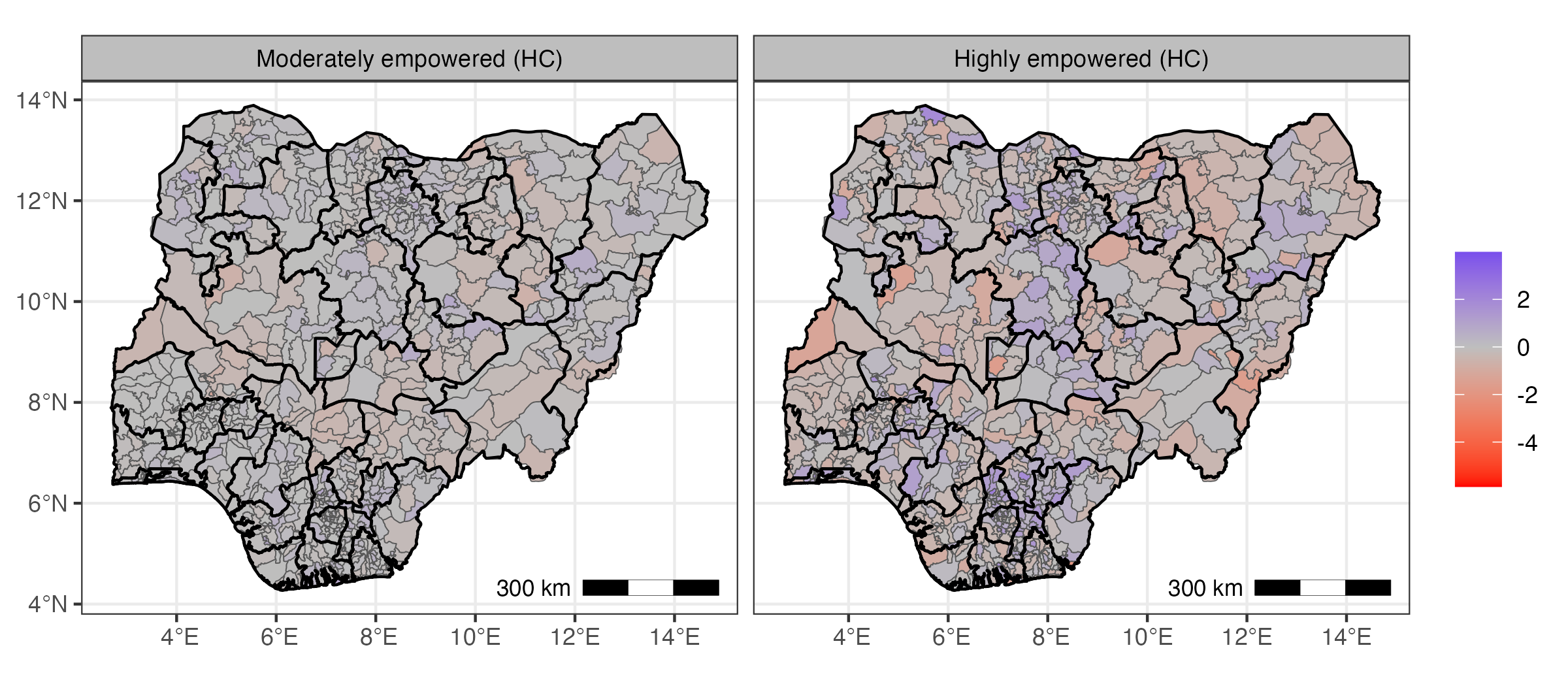}}
         {\includegraphics[scale=0.4]{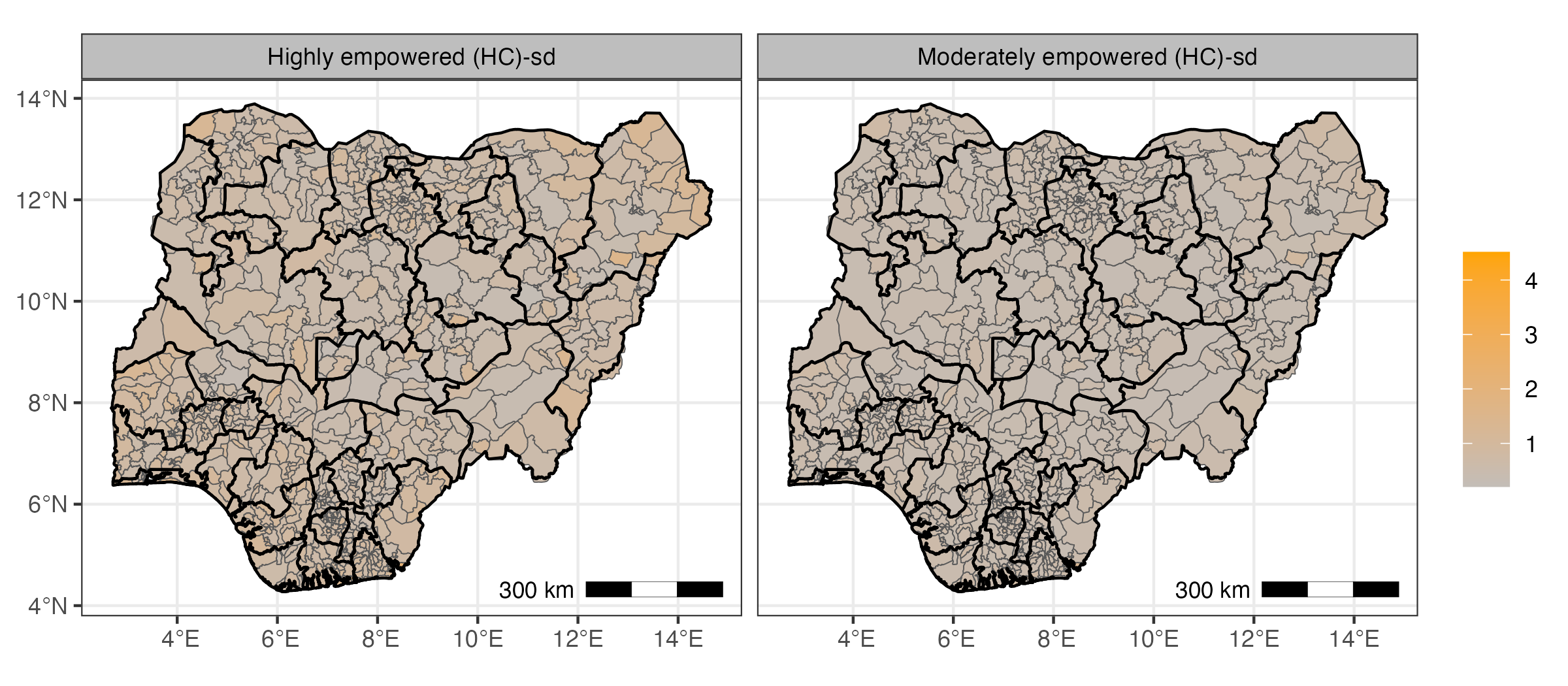}}\\
	 	{\includegraphics[scale=0.4]{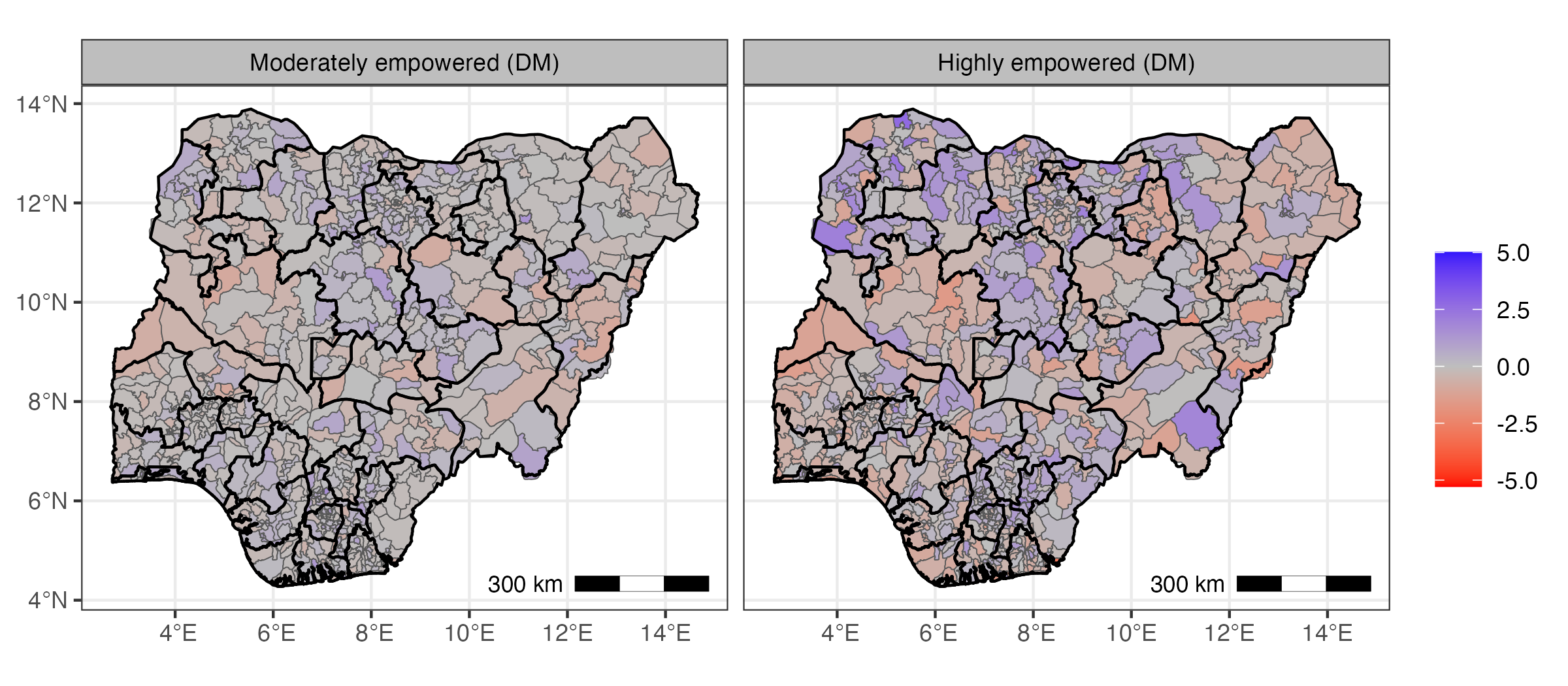}}
        {\includegraphics[scale=0.4]{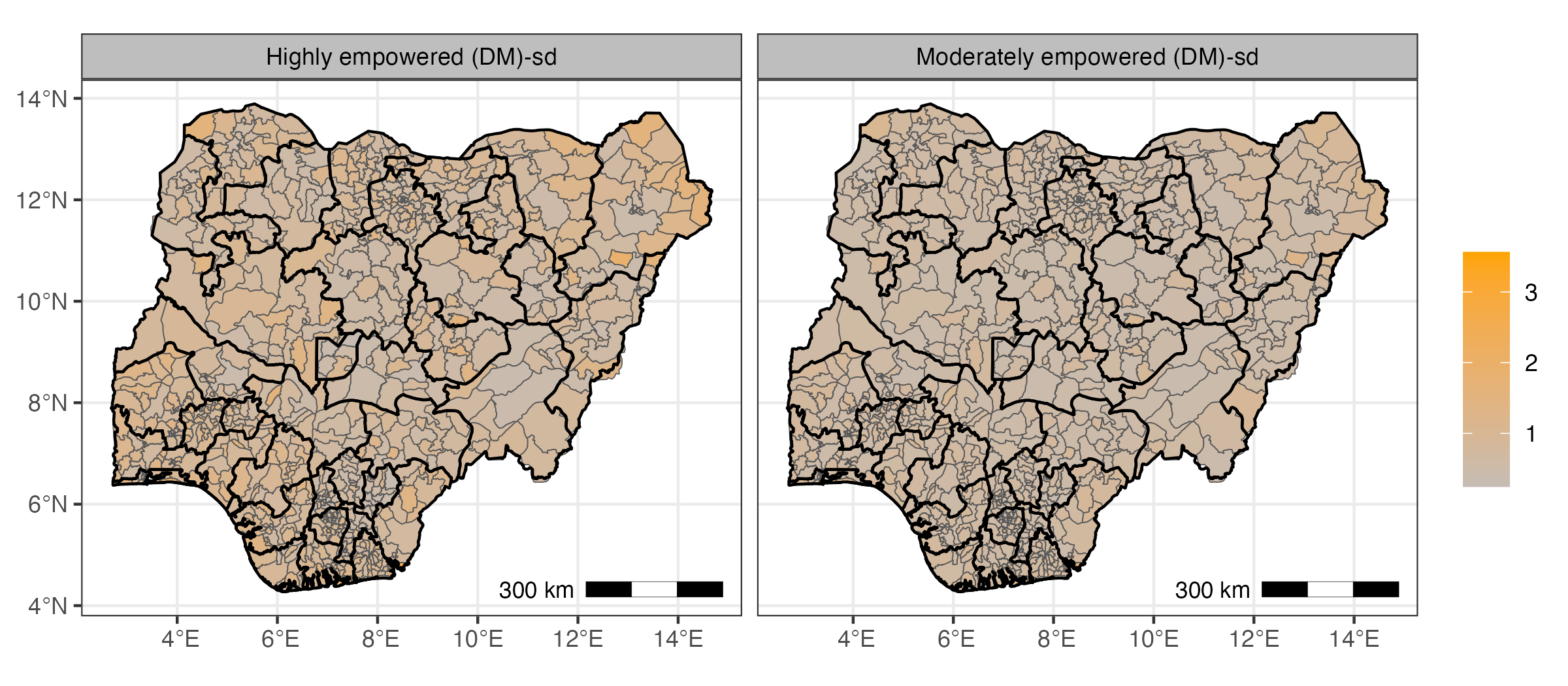}}
		\caption{\small Posterior mean and standard deviation of the effects $\hat \gamma$ of women's empowerment indicators on all basic vaccines coverage (defined as a child receiving a dose of BCG, three doses of DPT, three doses of oral polio vaccine (excluding the one at birth), and a dose of MCV). Data refer to children aged 12-23 months. }
  \label{eall}
\end{figure}

The estimates for the spatially varying effects of the two empowerment indicators are presented in Figure \ref{ezero} - \ref{eall}, where the upper panels show the maps for healthcare utilization while the lower panels are for decision-making. {Also included are the maps of the standard deviations, which convey the variability and precision of the model predictions across regions.} {In addition, Figures \ref{fig:va33} and \ref{fig:va44} in the Appendix show the posterior mean of the coverage probability conditional on different levels of empowerment.} Regarding zero-dose (Figure \ref{ezero}), maps for moderately empowered women show comparable moderate effects throughout the country and this is the case for both empowerment indicators. However, the findings shows wide variations among the highly empowered women with somewhat similar patterns for both empowerment indicators, where children of these women from most LGAs in the north-western part of the country, those from the south-east and a few other places in Taraba, Plateau, Adamawa and Kwara have lower chances of becoming zero-dose. Results for BCG (Figure \ref{ebcg}) show that women who are moderately empowered regarding healthcare utilization and living in most parts of Zamfara, Yobe, Gombe and Kwara are less likely to present their children for BCG vaccination while for most other parts of the country, the estimates are about zero, indicating minimal effects of being moderately empowered. Again, for both empowerment indicators, there are wide variations in the likelihood of receiving BCG vaccine among children of the highly empowered women. For healthcare utilization, children of the highly empowered women living in most part of Benue, FCT, Ekiti, and in the LGAs in the south-east and south-south regions of the country have higher chances of receiving the vaccine while for decision-making, children from many more LGAs including those in Kaduna, parts of Taraba, Adamawa, Kwara, and Kogi have higher chances of receiving BCG vaccine. 




Moving on to MCV-1 coverage (Figure \ref{emcv}), the chances of receiving the vaccine by children of the moderately empowered women are lower in only a few LGAs, based on both indicators. For the majority of the LGAs, the estimates are around zero, implying that in these places, being moderately empowered has minimal effect on the likelihood of receiving MCV-1 vaccine. However, for the highly empowered, particularly regarding healthcare utilization, there are many LGAs where the children are less likely to receive the vaccine. For participation in decision-making, children of the highly empowered women living in some neighboring LGAs in Adamawa and Taraba States, Kaduna, Kano, Benue, FCT, Ekiti, and across most parts of the south-east have higher chances of receiving the vaccine. The estimates for DPT-complete also show wide variations in the likelihood of receiving the complete dose among the highly empowered women where uptake is more likely among the children of women who are highly empowered regarding healthcare utilization and live in most states in the southern part of the country, in part of Kwara, Nasarawa, Plateau, southern Kaduna and the FCT. For participation in decision-making, the likelihood of receiving three doses of DPT vaccine are low except for a few LGAs scattered around the country. Lastly, for all basic vaccinations, variations in the estimates appear to be wider among the highly empowered women on healthcare utilization while for decision-making, the maps for both level of empowerment are somewhat similar, indicating high chances of receiving all basic vaccinations in only a few LGAs across the country.



\subsection{Validation results}
As described in Section \ref{val}, we validated the proposed model by computing and comparing the posterior predictive prevalence with the empirical prevalence per administrative state. That is, we predicted the prevalence from a higher spatial resolution (LGA) to a lower spatial resolution (states). Figure A3 in the supplementary material shows the plot of the posterior predictions $\hat p$ against the empirical prevalence across states for each vaccine indicator considered. Given that $\hat p$ and $\hat E$ approximately lie on the diagonal of the plot, it indicates that the proposed model is adequate, {since the correlation between $\hat p$ and $\hat E$ is high (0.87 for zerodose, 0.92 for DPT-complete, 0.82 for MCV1, 0.82 for BCG, and 0.94 for All basic)}. Figures A4 to A8 in the supplementary material 
show the spatial maps of these estimates,  $\hat p$ and $\hat E$ across years for all the vaccination indicators.


\section{Discussion}
As custodians of children, women have continued to play vital roles in vaccination uptake for their wards though in most developing countries, many of them are constrained due to the myriad of barriers they face that limit their capabilities. Thus, gender equity and women's empowerment are an integral part of the modern strategies for promoting healthy lives for all children. {While previous research has examined vaccination coverage in Nigeria, this study is the first to integrate a composite Women's Empowerment Index within a spatio-temporal framework to assess how empowerment interacts with local context and time-varying conditions to shape multiple dimensions of childhood immunization coverage across Nigeria.} 
We focused on participation in household decision-making and in utilization of healthcare facilities because of the potential to allow women to control and mobilize household resources for the well-being of their children, including the uptake of vaccination. 
We also quantify and predict the prevalence of these vaccine coverage at the LGA level in the country over time, allowing for the determination of the impact of previous interventions at different parts of the country. The findings indicate widespread variations in the use of vaccination services, particularly among the children of the highly empowered category of women, and this persists for all vaccination types. Furthermore, coverage for all basic vaccination was particularly low, around 10\% in most part of the country throughout the study period.

Previous studies have shown that when viewed at the national level or when multiple countries are combined, women's empowerment, particularly regarding participation in household decision-making and attitude toward wife beating, are significantly associated with immunization uptake \cite{amoah2023influence, singh2013maternal}. However, this study establishes the existence of 
huge variations across space in the modifying effects of empowerment regarding healthcare utilization and participation in household decision-making for all the vaccination schedules considered, with obvious variations in the case of the highly empowered women than it is for the moderately empowered. Although women with agency would more likely have control over resources and act without having to wait for their partner's approval, these women may also be challenged by time constraints due to competing out-of-home work schedule, limiting their ability to oversee their children's immunization. {In such contexts, empowerment alone may not translate into improved vaccination outcomes unless supportive health system structures, such as flexible clinic hours, mobile outreach programs and targeted health communication, and community childcare support are in place.} 
Where vaccination time conflicts with the time for income-generating activities, some caregivers may tend to delay vaccination \cite{babalola2009factors}, which {emphasizes the need for community-based solutions that reduce opportunity costs for working mothers.} 
{These findings showcase the importance of integrating women's empowerment initiatives with public health delivery strategies. Targeted interventions in areas such as Kaduna, Kebbi, Benue, and states in the South East, where empowerment correlates with lower rates of zero-dose children, could serve as models for leveraging empowerment in vaccine outreach. Strengthening community support networks and promoting flexible immunization schedules can ensure that empowerment effectively translates into health gains for children nationwide.}

Furthermore, while {
empowerment may be associated with higher vaccination uptake}, other factors beyond the control of women may cause local variations in access to vaccination services revealed by the maps. For example, the condition of access roads to an immunization center and travel time could serve as barriers, particularly for those in rural and urban slums. Separate studies in Kenya and Ethiopia \cite{okwaraji2012association,joseph2020spatial} found that travel times greater than one hour to a health facility lead to a substantial decline in immunization coverage. In Nigeria, Sato \cite{sato2020association} found that an increase in distance to the nearest health facility by 1 km reduces the probability of receiving any vaccine by approximately 5\%, delays the timing of vaccine uptake, but does not lead to dropout. Community engagement and education regarding child immunization could also motivate women to immunize their children, but these engagements may not happen at the same rate throughout the country, further causing variations in vaccine coverage \cite{babalola2009factors}. On the supply side, service delivery, adequate and regularity of vaccine supplies are critical factors for immunization, but vaccine stockouts remain a major challenge across the country, particularly at the LGA level \cite{etim2024everything,gooding2019impact}. 
Studies by Eboreime et. al \cite{eboreime2015access} reported state-level variations in human resources and commodity supply at routine immunization service delivery points in Nigeria, and that in general, residents of northern Nigeria tend to have a service delivery point within a radius of 5 km from their settlements, and conversely, it is substantially lower in the southern part. However, vaccination uptake is generally lower in the north, a condition that could be explained by socio-cultural disparities in immunization uptake across the country.  { These findings may suggest that increasing women's empowerment alone is insufficient to eliminate inequalities in immunization uptake. Structural and contextual barriers, such as poor road networks, long travel times, vaccine stock-outs, and shortages of skilled health personnel, must be addressed simultaneously. Policies should prioritize improving geographic accessibility to immunization services by expanding community outreach programs, mobile vaccination units, and transportation support, particularly in rural and hard-to-reach areas.}

Women who are moderately or highly empowered and live in the northern part of the country appear to be less likely to present their children for BCG vaccines, like most of the other vaccines. The Nigeria national immunization guidelines recommend that BCG vaccines be administered at birth, preferably within 24 hours or as soon as possible \cite{unicef2024}. Regular attendance in antenatal care services, institutional delivery,  and knowledge of immunization schedule have been identified to encourage birth-dose immunization in the country \cite{ibraheem2019determinants} although studies by Olakunle and colleagues \cite{olakunde2022coverage} found that children born in private health facilities are less likely to receive this vaccine. However, institutional deliveries and antenatal care attendance are less common in northern Nigeria when compared with the southern part of the country \citep{gayawan2014,egbon2024modeling}. In the North-East and North-West regions of Nigeria, vaccine distrust, misconception, religious norms, and spousal and family disapproval were reported to hinder vaccine uptake \cite{etim2024everything}. {Policies should prioritize linking immunization with antenatal and delivery services, expanding health facility-based births, and addressing vaccine mistrust through community and religious engagement. Promoting vaccine education and male involvement can help translate women's empowerment into improved vaccination coverage and reduce regional disparities.}

The predicted prevalence maps show heterogeneity over space and time for all the vaccine types, and that the prevalence of all basic vaccinations is particularly low throughout the country. Over the years, Nigeria has aimed to scale up targeted immunization coverage and bridge inequality, and therefore implemented different interventions to strengthen both the supply and demand sides of immunization. For example, following a public health concern on routine immunization across the country, a National Emergency Routine Immunization Coordination Centre (NERICC) was established in 2017 with the mandate of improving detection and responsiveness in the resolution of routine immunization gaps, among others. Other efforts include strengthening Primary Health Care Under One Roof, a national policy aimed at improving Primary Health Care performance to contribute to better health outcomes; and the implementation of the zero-dose reduction plan through routine immunization and supplementary immunization activities \cite{jean2024high}. The maps generated in this study provide an opportunity to evaluate the impact of these and other interventions that have been implemented over the years. 

It is worth mentioning some limitations of the study. The study relies on survey data, which may not provide a true representation of vaccine coverage, particularly in difficult-to-reach and conflict-prone areas. These areas may harbor a high number of unvaccinated children. For instance, during the 2018 survey, some LGAs in Borno state were specifically left out due to conflict. The data were self-reported by the caregivers, and thus the immunization data can be inflated due to a number of reasons, for instance fear of being reprimanded or being penalized for not immunizing the child. Moreover, the sample data are not quite representative at the LGA level, as there are those without any sample, although the spatial technique adopts some mechanism to borrow strength across neighbouring locations to improve local estimates. Furthermore, the cross-sectional nature of the data limits the ability to make causal inferences from the study. {Lastly, a multivariate model could be valuable for exploring shared spatial structures across vaccines and allow for cross-correlation among vaccines to be measured. We view this as an important direction for future work.} Nonetheless, the geospatial methodology allows the use of computing technology to provide insight into an aspect of the geography of immunization coverage in Nigeria, and the findings could help identify places where efforts to empower women could positively impact vaccine uptake.

 \section{Conclusion}
Women have continued to play an important role in ensuring the well-being of their children, but gender disparity and women's disempowerment have remained as major barriers to healthcare utilization, including uptake of immunization services in most LMIC. This is particularly the case for Nigeria, where gender segregation provides material advantages for men while limiting the role of women, particularly in decision-making processes. Thus, understanding the spatially varying patterns of aspects of women's empowerment and immunization uptake provides opportunities to glean the spatial overlap and ultimately guide the adoption of newer and more effective interventions to increase immunization coverage in Nigeria. Our findings highlight that highly empowered women are less likely to have zero-dose children, and participation in decision-making increases the chances of coverage of the MCV-1 vaccine. However, inequality in immunization uptake appears to be pronounced among the highly empowered women, pointing to the fact that engagements in other economic or social activities might conflict with vaccination time. Deliberate efforts might therefore be required to enhance the empowerment capabilities of women, particularly in northern Nigeria, and at the same time to initiate efforts that ensure that these women do not relegate vaccination periods of their wards.

\newpage

\subsection*{Acknowledgements}
We would like to express our gratitude to Katy Gaythorpe for reading the manuscript and providing us with valuable comments that helped in improving on the earlier draft. We also thank Jonathan Mosser and Emily Haeuser for their comments on the results.

\subsection*{Authors' contributions}
EG conceived the idea, participated in data curation, formal analysis, and writing of manuscript. OAE participated in data curation, formal analysis and writing of manuscript. EU participated in formal analysis and drafting of manuscript. JAU participated in the writing of manuscript and CT supervised the project.
\subsection*{Funding}
This work was carried out as part of the Vaccine Impact Modelling Consortium (www.vaccineimpact.org), but the views expressed are those of the authors and not necessarily those of the Consortium or its funders. The funders were given the opportunity to review this paper prior to publication, but the final decision on the content of the publication was taken by the authors.

 This work was supported, in whole or in part, by the Bill \& Melinda Gates Foundation, via the Vaccine Impact Modelling Consortium [Grant Number INV-034281], previously (OPP1157270/INV-009125), the Wellcome Trust via the Vaccine Impact Modelling Consortium [Grant Number 226727\_Z\_22\_Z], and Gavi, the Vaccine Alliance. Under the grant conditions of the Foundation, a Creative Commons Attribution 4.0 Generic License has already been assigned to the Author Accepted Manuscript version that might arise from this submission.
\newpage

\appendix
\section*{Appendix}
\setcounter{figure}{0} \renewcommand{\thefigure}{A.\arabic{figure}} 
\setcounter{table}{0} \renewcommand{\thetable}{A.\arabic{table}} 

\begin{figure}[H]
    \centering
    \includegraphics[scale=0.7]{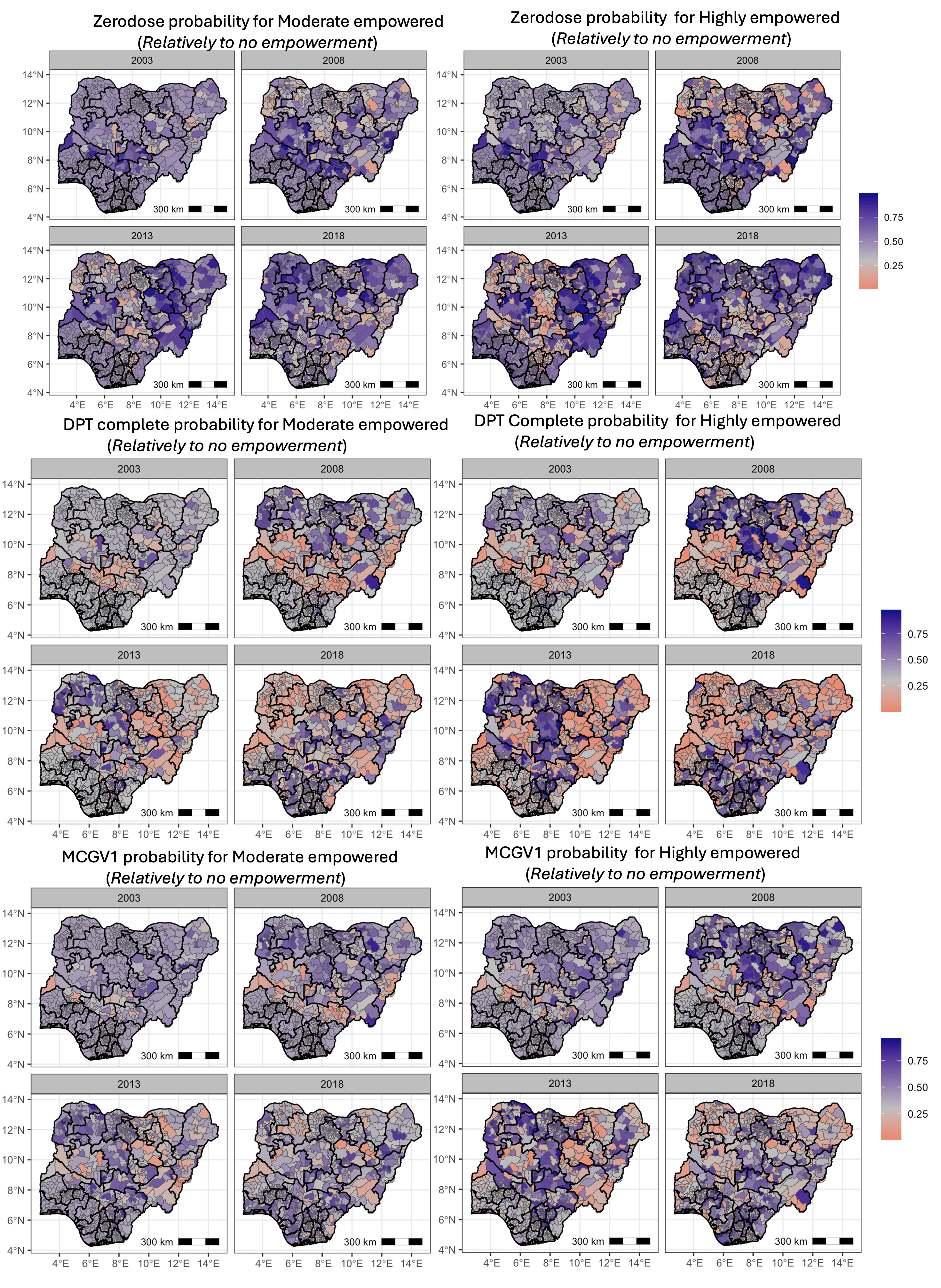}
    \caption{Posterior predicted mean of vaccination coverage conditional on empowerment level for Zerodose, DPT complete, and MCGV1 vaccines.}
    \label{fig:va33}
\end{figure}
\begin{figure}[H]
    \centering
   \includegraphics[scale=0.65]{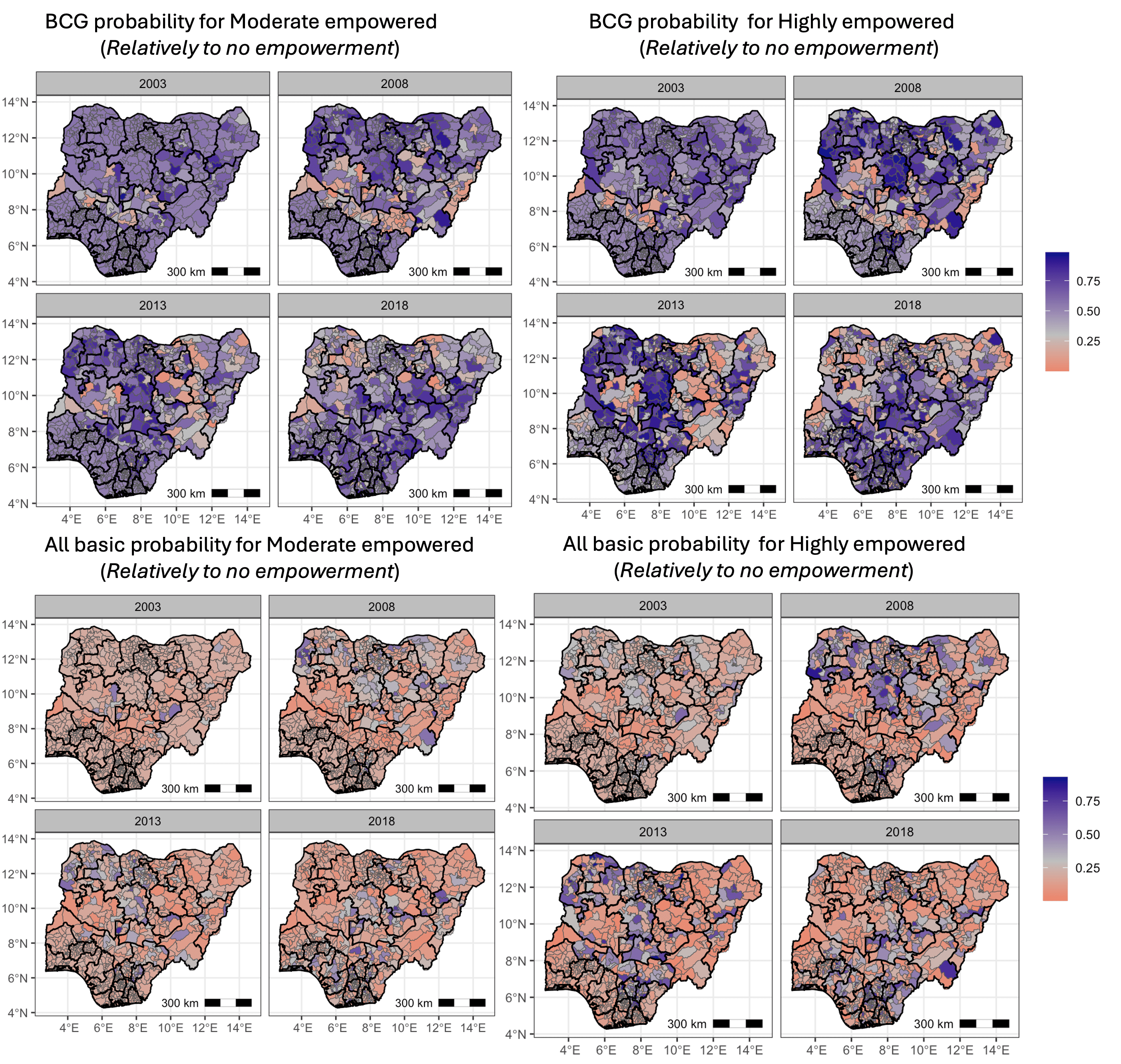}
    \caption{Posterior predicted mean of vaccination coverage conditional on empowerment level for BCG and Allbasic vaccines.}
    \label{fig:va44}
\end{figure}
\end{document}